\documentclass[pdflatex,sn-mathphys-num]{sn-jnl}

\usepackage{graphicx}%
\usepackage{multirow}%
\usepackage{amsmath,amssymb,amsfonts}%
\usepackage{amsthm}%
\usepackage{mathrsfs}%
\usepackage[title]{appendix}%
\usepackage{xcolor}%
\usepackage{textcomp}%
\usepackage{manyfoot}%
\usepackage{booktabs}%
\usepackage{algorithm}%
\usepackage{algorithmicx}%
\usepackage{algpseudocode}%
\usepackage{listings}%
\usepackage{braket}
\usepackage{relsize}
\usepackage{soul}
\usepackage{multirow}  
\usepackage{booktabs}  
\usepackage{array}     

\theoremstyle{thmstyleone}%
\theoremstyle{thmstyletwo}%

\theoremstyle{thmstylethree}%

\begin{document}


\title[Article Title]{Tunable polarization-entangled near-infrared photons from orthogonal GaAs nanowires}


\author*[1]{\fnm{Elise} \sur{Bailly-Rioufreyt}}\email{ebailly@phys.ethz.ch}

\author[1]{\fnm{Zoya} \sur{Polshchykova}}

\author[1]{\fnm{Grégoire} \sur{Saerens}}

\author[1]{\fnm{Wenhe} \sur{Jia $^{\dagger}$}}

\author[2]{\fnm{Thomas} \sur{Dursap $^{\ddagger}$}}

\author[1]{\fnm{Andreas} \sur{Maeder}}

\author[2]{\fnm{Philippe} \sur{Regreny}}

\author[1]{\fnm{Robert J.} \sur{Chapman}}

\author[1]{\fnm{Helena} \sur{Weigand}}

\author[2]{\fnm{Alexandre} \sur{Danescu}}

\author[2]{\fnm{Nicolas} \sur{Chauvin}}

\author[2]{\fnm{José} \sur{Penuelas}}

\author[1]{\fnm{Rachel} \sur{Grange}}


\affil[1]{\orgdiv{\normalsize{ETH Zurich, Department of Physics}}, \orgname{\normalsize{Institute for Quantum Electronics, Optical Nanomaterial Group}}, \orgaddress{\street{\normalsize{Auguste-Piccard-Hof 1}}, \city{\normalsize{Zurich}}, \postcode{8093}, \country{Switzerland}}}


\affil[2]{\orgname{\normalsize{CNRS, ECL, INSA Lyon, UCBL, CPE Lyon, INL, UMR 5270}}, \orgaddress{ \city{\normalsize{Ecully}}, \postcode{\normalsize{69130}}, \country{\normalsize{France}}}}

\affil[$\dagger$]{Present address : \orgdiv{\normalsize{Department of Electrical and Computer Engineering}}, \orgname{\normalsize{National University of Singapore}}, \orgaddress{\postcode{\normalsize{117583}}, \country{\normalsize{Singapore}}}}

\affil[$\ddagger$]{Present address : \orgname{\normalsize{IMEC, Kapeldreed 75}}, \orgaddress{ \city{\normalsize{Leuven}}, \postcode{\normalsize{3001}}, \country{\normalsize{Belgium}}}}


\abstract{Quantum entanglement is a fundamental resource for emerging quantum technologies, enabling secure communication and enhanced sensing. For decades, generating polarization entangled states has been mainly achieved using bulk crystals with spontaneous parametric down conversion (SPDC), preventing scalability and on-chip integration. Miniaturizing the quantum source provides access to more versatility and tunability while enabling an easier integration to other devices, notably necessary for satellite-based quantum communication, and eventually reducing fabrication costs. This challenging task can be achieved with Zinc Blende GaAs nanowires. They already have shown an efficient photon pairs generation via SPDC at 1550 nm. Here we demonstrate that a pair of orthogonal GaAs nanowires constitutes a new nanoscale platform to control the quantum state at telecommunication wavelength, enabling a transition from polarization entangled to separable states as a function of the pump polarization, with fidelities reaching $\text{90}\%$.}


\maketitle

\section{Introduction}\label{sec1}
Entanglement, and more specifically polarization entangled photons, are at the core of quantum applications, ranging from quantum key distribution and quantum communication \cite{curty_entanglement_2004,scarani_security_2009,erven_entangled_2008,bouwmeester_experimental_1997}, to quantum imaging and sensing \cite{moreau_qimaging_2019,defienne_polarization_2021}. Among the methods to generate polarization entanglement, spontaneous parametric down conversion (SPDC) has been the prevalent process, for which a non-centrosymmetric crystal generates signal and idler photon pairs of frequencies given by energy conservation \cite{Burnham70}. This method has the particular advantage that the emission directions of the signal and idler photons are well-correlated. In most platforms, and especially for free space applications, bulk crystals have been used to generate entanglement with SPDC \cite{kwiat_new_1995,kwiat_ultrabright_1999,Edamatsu2007}, preventing their integration on a chip and requiring the fulfillment of restrictive phase matching conditions. \\
\indent Recent years have seen considerable effort in miniaturizing quantum sources for free space applications. The development of SPDC in thin films \cite{okoth_microscale_2019,santiago-cruz_entangled_2021} for which phase matching conditions are relaxed \cite{okoth_idealized_2020}, has opened the path to “flat” quantum optics. In order to counterbalance the low interaction length in thin films (and thus the low efficiency), two methods in particular have been explored. The first consists in developing metasurfaces that generate polarization entanglement \cite{ma_polarization_2023,jia_polarization-entangled_2025} and enhance light matter interaction \cite{noh_fano_2025,zhang_spatially_2022,weissflog_directionally_2024,santiago-cruz_photon_2021,fan_enhanced_2025}. This method requires a careful design of the metasurface in order to have a spatial overlap of the electromagnetic modes \cite{gigli_quasinormal-mode_2020} and is commonly achieved through demanding top-down fabrication. The second method exploits materials with exceptionally high second-order susceptibility tensor components $\chi^{(2)}$, such as GaP \cite{santiago-cruz_entangled_2021}, also used to generate polarization-entanglement upon post-selection \cite{sultanov_flat-optics_2022}, or 2D materials \cite{guo_ultrathin_2023,feng_polarization-entangled_2024, liang_tunable_2025,weissflog_tunable_2024,Guo24,Kallioniemi25} where their intrinsic $\chi^{(2)}$ properties are exploited to generate entanglement with high coincidence rates. However, achieving large-area synthesis of 2D materials with high scalability and free of imperfections is a major challenge \cite{abbas_recent_2024}. \\
\indent GaAs nanowires (NWs) are a new promising platform, both agile and efficient, for quantum applications at the nanoscale. Not only do GaAs NWs have an intrinsically high $\chi^{(2)}$, which can reach 370 pm/V \cite{boyd2008nonlinear,shoji_absolute_1997}, but due to their bottom-up fabrication, they also present key advantages over conventional top-down approaches. They exhibit a high crystalline quality with minimal defects, while the growth parameters can be finely tuned to control their diameter and length. Furthermore, bottom-up approaches enable the engineering of complex heterostructures that can serve as building blocks for multifunctional devices \cite{Barrigon19,Hyun13,Dalacu19}. Additionally, NWs can be synthesized on a wide variety of substrates, making them suitable for integration into unconventional platforms. Once fabricated, they can also be transferred to other substrates and easily manipulated using atomic force microscope tips or micromanipulators \cite{Conache08,Junno95}. In particular, it has recently been shown that GaAs NWs were mechanically transferred onto adequate transparent substrates (without fluorescence or nonlinear contributions) after their epitaxial growth, leading to background-free photon pairs emitted via SPDC \cite{saerens_background-free_2023}. \\
\indent Here, we show that we can exploit type-0 SPDC in GaAs NWs to generate polarization entanglement by pumping two perpendicularly arranged NWs with a linearly polarized laser. In type-0 SPDC, the pump, signal, and idler photons all share the same polarization. By tuning the pump polarization, we are able to generate Bell states, maximally entangled when the pump polarization is approximately at $\pm45^{\circ}$ of the nanowire axes due to the superposition of the two photon pairs emitted by each NW, using a known method for generating polarization entanglement \cite{kwiat_ultrabright_1999,Guo24,Kallioniemi25}. With the same nanowires, separable states are achieved when the pump polarization aligns with the respective axis of the nanowires. We thus demonstrate a fully tunable quantum source capable of generating entangled-to-separable states from two orthogonal nanowires, without any post-selection. Furthermore, this quantum nanosource operates at telecom wavelengths, a range highly sought after due to its ability to enable long-distance, low-loss transmission, its compatibility with existing optical fiber networks and its mature low-noise technology, making them adequate for quantum communication and networking applications \cite{Gisin07,Pittaluga25}.

\section{Results}\label{sec2}
\subsection{Characterization of the effective nonlinear susceptibility tensor}

\begin{figure}[h!]
   \centering
    \includegraphics[width = 0.7\textwidth]{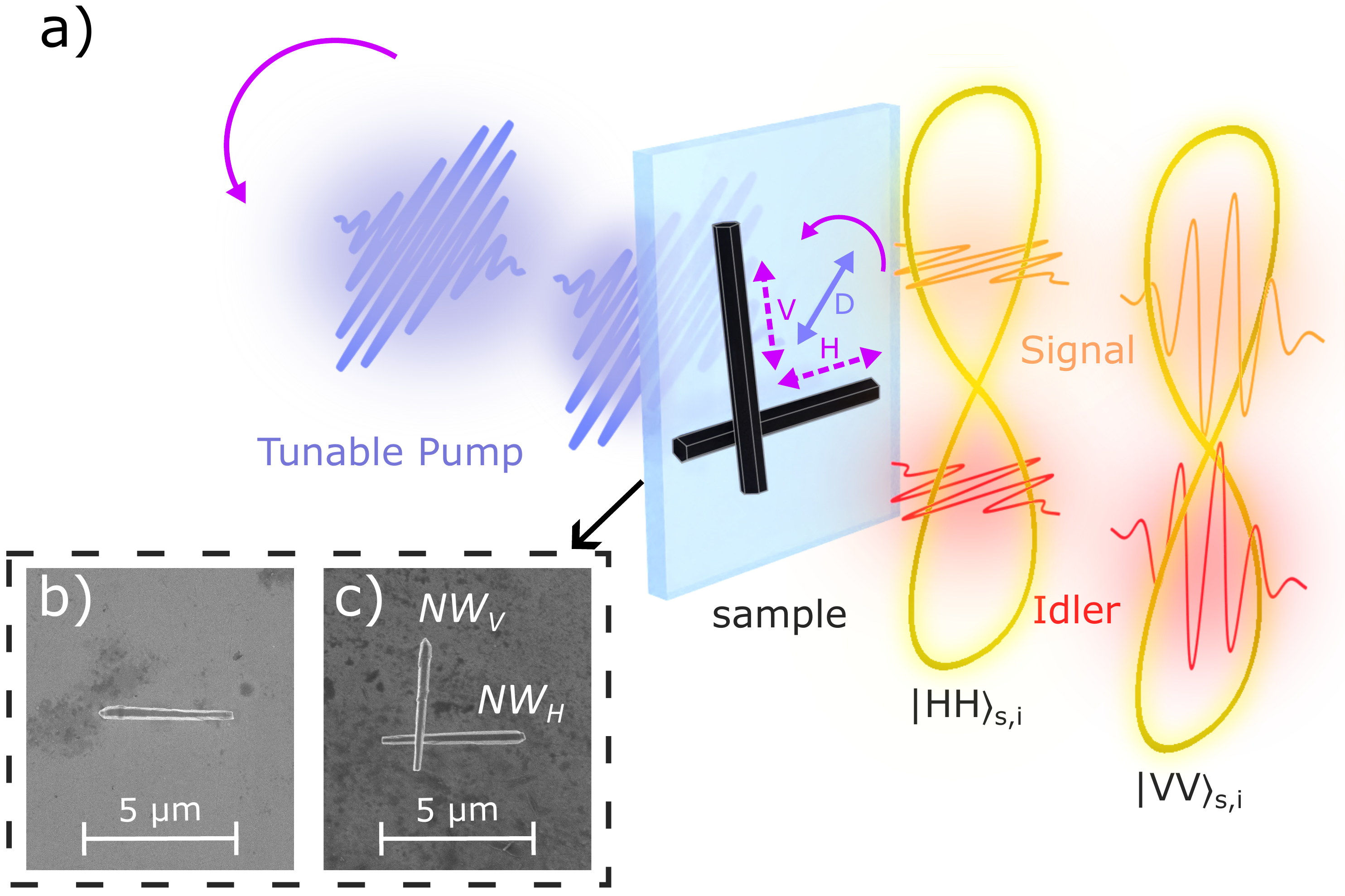}
  \caption{GaAs nanowires for the generation of tunable polarization entangled photon pairs. a) Sketch of the entangled idler and signal photons emitted from two orthogonal GaAs nanowires when pumped by a diagonal polarization (represented by the blue arrow). The pump polarization can also be rotated, as shown with the pink arrows, to probe selectively one single NW with a horizontal (H) or vertical (V) pump polarization, generating separable states. b) SEM image of a single GaAs used to characterize the susceptibility tensor of the nanowire. c) SEM image of the tunable nanosource consisting of two orthogonal nanowires labelled $\mathrm{NW_{H}}$ and $\mathrm{NW_{V}}$.}
  \label{Fig1}
\end{figure}

Single Zinc Blende (ZB) GaAs NWs have previously been used for the generation of background-free photon pairs from type-0 SDPC with high coincidence rates, up to 60 GHz/Wm \cite{saerens_background-free_2023}. Here, the nanosource of polarization entangled photons consists of two orthogonal ZB GaAs NWs grown in the (111) direction, laid on a transparent $\mathrm{SiO_{2}}$ substrate, as shown in the schematic diagram in Fig. \ref{Fig1}a.

\begin{figure}[h!]
   \centering
    \includegraphics[width = \textwidth]{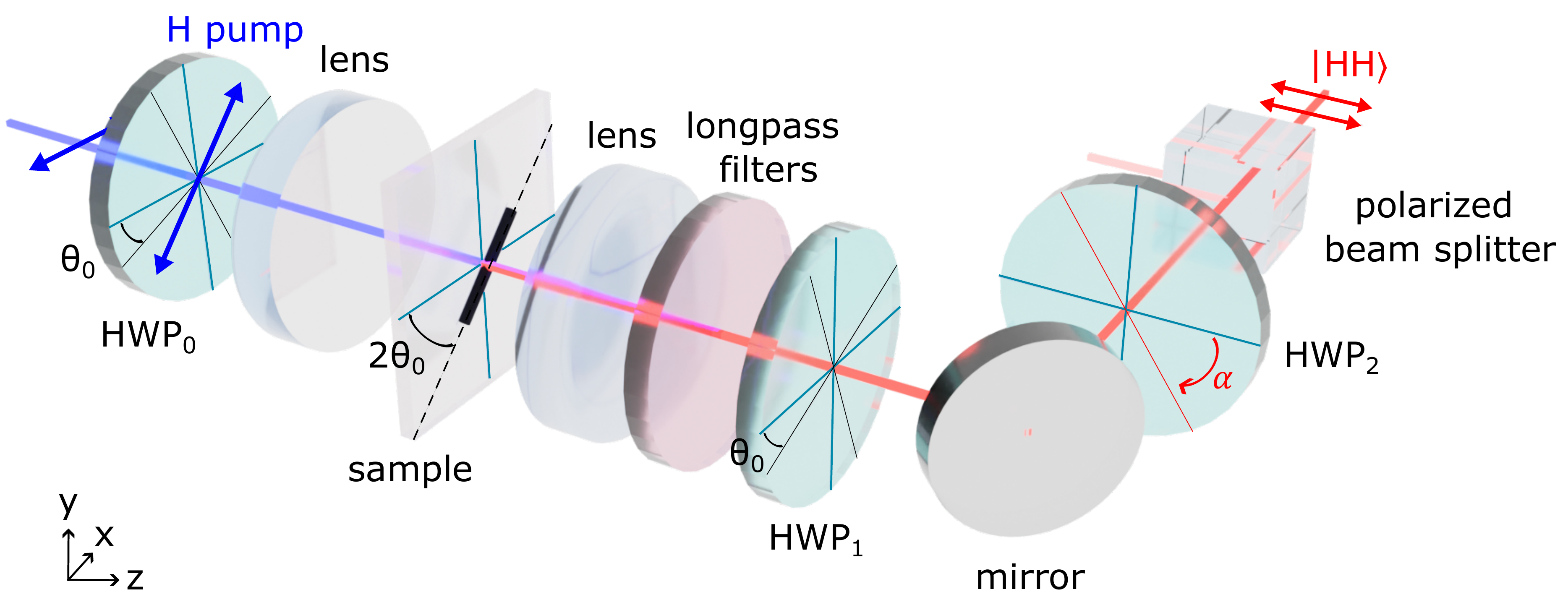}
  \caption{Characterization of a single NW. A lens (f = 8 mm) focuses a continuous wave (CW) pump beam (at 778 nm) on the nanowire. The SPDC generated photons are collected with an identical lens, filtered by longpass filters. The half wave plate $ \mathrm{HWP_{0}}$ controls the pump polarization, $ \mathrm{HWP_{1}}$ rotates the photons emitted along the long axis of the nanowire on the horizontal x-axis while $ \mathrm{HWP_{2}}$ rotates the polarization of the quantum state. The polarized beam splitter (PBS) transmits only the horizontal polarization. The idler and signal photons are then coupled to a fiber, separated by a fiber splitter and detected by two superconducting nanowire single-photon detectors (SNSPDs). (More details in the Supplementary Information.)}
  \label{Fig2}
\end{figure}

The NWs were grown by molecular beam epitaxy and mechanically transferred to a glass substrate (see Methods for more details). An SEM image of the NWs is shown in Fig. \ref{Fig1}c). Both NWs are illuminated by a linearly polarized pump laser and the resulting signal is a superposition of the photon pairs generated by each NW via SPDC. This mechanism relies on second order nonlinear processes, which are governed by the interaction between the electric fields of light and the nonlinear polarization encoded in the second order susceptibility tensor $\chi^{(2)}.$ In particular, the SPDC polarization response can be inferred through the quantum-classical correspondence between SPDC and sum-frequency generation \cite{Poddubny16}. However, the susceptibility tensor of GaAs NWs can differ from the bulk ZB crystal due to surface effects arising from the large surface-to-volume ratio \cite{Guyot86,Zhang13,He13}, polytypism \cite{Pimenta16,Timofeeva16} or the presence of defects \cite{Zhang21,Schaller02}. It has also been shown that the crystal orientation in ZB GaAs NWs can flip across a twin plane while maintaining overall crystallinity \cite{Dursap25}. Moreover, the NW can also support electromagnetic modes \cite{DeCeglia19}, which locally change the fields strengths. These considerations motivate the experimental determination of the effective susceptibility tensor of a GaAs NW through projection measurements. To this end, we use a single NW as a reference (Fig. \ref{Fig1}b), and characterize it with the setup of Fig. \ref{Fig2}. 

\begin{figure}[h!]
   \centering
    \includegraphics[width =0.9 \textwidth]{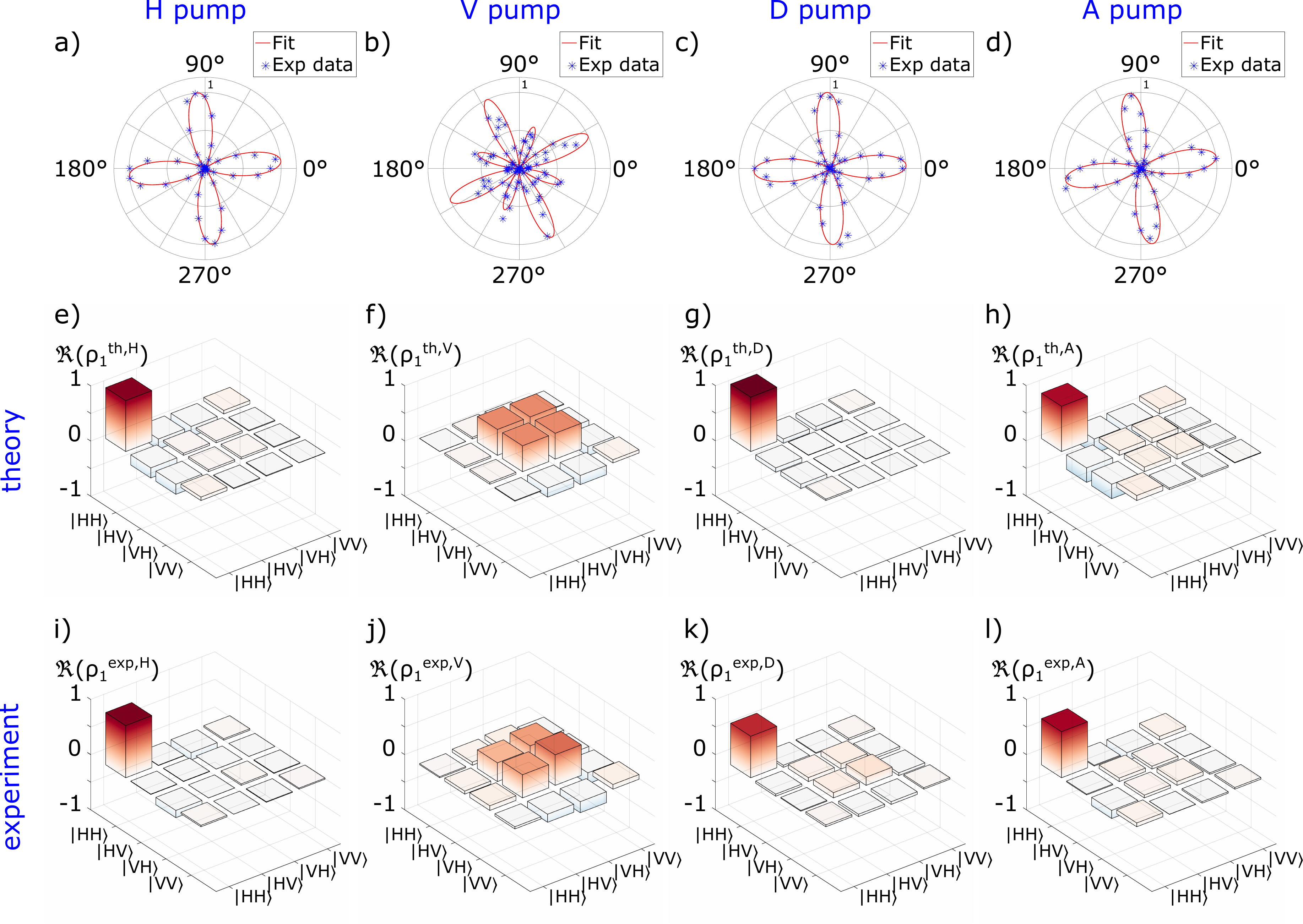}
  \caption{Characterization of a single GaAs nanowire. a)-d) Normalized experimental coincidence counts as a function of the angle of rotation of $\mathrm{HWP_{2}}$ (blue stars), and normalized theoretical coincidence counts obtained with Eq. \ref{Eq1} after minimization of the cost function (see Methods), in red lines. Unnormalized experimental  coincidence counts, shown in the Supplemenray Information, reveal that photon coincidences are about an order of magnitude higher under H pump polarization compared to V ones. e-h) Real part of the theoretical density matrix $\rho_{\mathrm{1}}^{\mathrm{th,P}} =\ket{\Psi_{\mathrm{1}}^{\mathrm{th,P}}}\bra{\Psi_{\mathrm{1}}^{\mathrm{th,P}}} $, with P the pump polarization. i-l) Real part of the experimental density matrix $\rho_{\mathrm{1}}^{\mathrm{exp,P}}$. Each column correspond to a given P pump polarization being horizontal (H) for a),e),i); vertical (V) for b),f),j); diagonal (D) for c),g),k); and antidiagonal (A) for d),h),l).} 
  \label{Fig3}
\end{figure}

A continuous-wave (CW) linearly polarized laser at 778 nm and with a power of 12 mW is focused on the sample with a spot size of 4 \textmu m (at full width at half maximum). To determine the pump polarization, we rely on prior observations that SPDC in similar NWs occurs with a type-0 process, reaching maximum photon-pair generation when the pump polarization is parallel to the NW's long axis \cite{saerens_background-free_2023}. Similarly, the intensity of the Second Harmonic Generation (SHG) is maximized for this specific pump polarization. Thus, we experimentally identify the long axis orientation of the NW by rotating the angle $\theta$ of a first half wave plate (HWP), labelled $\mathrm{HWP_{0}}$ in Fig \ref{Fig2}a, while monitoring the intensity of the SHG onto a camera when the NW is illuminated by a femtosecond laser at 1555 nm. The SHG intensity reaches its maximum at $\theta = \theta_{0}$. The pump polarization associated to this specific orientation, parallel to the NW long axis, will be referred as H polarization. A second HWP (labelled $\mathrm{HWP_{1}}$) is then inserted in the detection path with the same angle $\theta_{0}$ to project the photons emitted along the NW long axis onto the horizontal x-axis of the lab frame. Finally, a third HWP (labelled $\mathrm{HWP_{2}}$) is inserted to rotate the polarization of the emitted photons as a function of its rotation angle $\alpha$. The photons are then projected on the horizontal x-axis with a polarized beam splitter (PBS) before being coupled to a fiber, probabilistically split via a fiber splitter and detected with superconducting nanowire single-photon detectors. The  coincidence counts are then recorded as a function of $\alpha$ and as a function of the pump polarization (Fig. \ref{Fig2} and \ref{Fig3}). The latter is varied among four distinct orientations. The first one is chosen along the NW long axis, for which $\theta^{H} = \theta_{0}$. As mentioned earlier, since $\mathrm{HWP_{1}}$ is set to project this specific axis on the horizontal x-axis, we will refer to this  orientation as the horizontal (H) pump polarization (Fig \ref{Fig3}a). We also set the pump polarization orthogonal to the NW long axis (with the angle of $\mathrm{HWP_{0}}$ being $\theta^{V} = \theta_{0} +45^{\circ}$, corresponding to the vertical (V) pump polarization - Fig \ref{Fig3}b), and at $\pm45^{\circ}$ of the long axis ($\theta^{D/A} =\theta_{0} \pm22.5^{\circ}$, corresponding to the diagonal (D) and antidiagonal (A) pump polarizations - Fig \ref{Fig3}c,d). The normalized polar plots of the experimental coincidence counts are shown as the function of the angle $\alpha$ in blue dots in Fig. \ref{Fig3}a-d). Since $\mathrm{HWP_{2}}$ rotates the polarization vector by an angle $2\alpha$, the data points from 180$^{\circ}$ to 360$^{\circ}$ should be a replica of the data plotted from 0$^{\circ}$ to 180$^{\circ}$, as confirmed experimentally. These additional data points enable to check the stability of the signal and enhance the statistical robustness for the fitted estimation of the effective $\chi^{(2)}$, as shown later. The coincidence counts recorded for the H pump polarization (Fig. \ref{Fig3}a) are maximal when $\alpha$ is a multiple of $90^{\circ}$, which confirms a type-0 process, with the photons being polarized as well along H. Equivalently, this means that the $\chi^{(2)}_{xxx}$ plays a more significant role than the $\chi^{(2)}_{xij}$, with $i,j \in \{x,y\}$.
On the other hand, the coincidence counts obtained when the NW is pumped along V (Fig. \ref{Fig3}b) suggest a more complex emission mechanism that needs to include several components of the $\chi^{(2)}$ tensor. The coincidence counts measured for D and A pump polarizations suggest that the emitted photons are mainly polarized along the H axis, with a negligible contribution along the V axis. In other words, the $\chi^{(2)}_{xxx}$ component should be larger than the $\chi^{(2)}_{yij}$ with $i,j \in \{x,y\}$. The unnormalized coincidence counts shown in the Supplementary Information reveal that the coincidence count rate is indeed an order of magnitude higher for the H pump compared to the V pump.\\
\indent To experimentally fully characterize our two photon state generated from a single NW, the coincidence-to-accidental ratio (CAR) was also measured as a function of the power pump, under H pump polarization, and for $\alpha=0^{\circ}$. The CAR reaches 50 at 12 mW and exceeds 150 for pump powers below 2 mW, demonstrating performances comparable to the state of the art for quantum nanosources \cite{liang_tunable_2025, Guo24} (see Supplementary Information). \\

In order to extract the susceptibility tensor values from the experimental projection measurements, we have to theoretically predict the number of coincidence counts. As shown in the Methods section, the coincidence counts are obtained by projecting the quantum state on the horizontal axis. The quantum state measured at the detectors after propagation can be expressed not only as a function of the Jones matrices of the optical detection components but also as a function of the quantum state generated from a single NW by SPDC, labelled $\ket{\Psi_{\mathrm{1} }^{\mathrm{th, P}}}$, which can be defined as 
\begin{equation}
     \ket{\Psi_{\mathrm{1} }^{\mathrm{th, P}}}(\mathbf{u}) = \sum_{i\in \{x,y\}}E_{i}^{\mathrm{P}}\mathbf{u}_{i}.
     \label{Eq1}
\end{equation}

\noindent This compact notation includes the pump polarization $\mathrm{P}$ (with $\mathrm{P}\in$\{H, V, D, A\}), the normalized pump field amplitude  $E_{i}^{\mathrm{P}}$ (see Methods), as well as the 
material contribution with  $\mathbf{u} =[\mathbf{u}_{x} ~ \mathbf{u}_{y}]$ being related to the susceptibility tensor via :

\begin{equation}
    \mathbf{u}_{i} =
     \begin{bmatrix}
u_{i1} \\
u_{i2} \\
u_{i3} \\
u_{i4} 
\end{bmatrix} \propto
     \begin{bmatrix}
\chi^{(2)}_{\mathrm{ixx}} \\
\chi^{(2)}_{\mathrm{ixy}} \\
\chi^{(2)}_{\mathrm{iyx}} \\
\chi^{(2)}_{\mathrm{iyy}}
\end{bmatrix}
\label{Eq2}
\end{equation}
in the \{HH, HV, VH, VV\} basis with $\chi^{(2)}_{\mathrm{ijk}}$ the nonlinear susceptibility tensor values of the GaAs NW. As previously discussed, these values are expected to differ from the ZB ones, motivating the estimation of the effective tensor. 
To that end, we define a $\{\mathbf{u}\}$-dependent cost function $f_{\mathrm{cost}}$, that computes the difference between the experimental and theoretical coincidence counts, normalized by the respective maximal values, as detailed in the Methods section. The minimization of the cost function $f_{\mathrm{cost}}$ leads to an estimation of the effective susceptibility tensor $\chi^{(2)}_{\mathrm{eff}}$:

\begin{equation}
\chi^{(2)} _{eff}\propto
\begin{bmatrix}
1.00 & 0.069 & \mathsmaller{\mathrm{NA}}  & \mathsmaller{\mathrm{NA}}  & \mathsmaller{\mathrm{NA}}  & -0.198 \\
0.008 & -0.042 & \mathsmaller{\mathrm{NA}}  & \mathsmaller{\mathrm{NA}}  & \mathsmaller{\mathrm{NA}}  &0.119\\
\mathsmaller{\mathrm{NA}}  & \mathsmaller{\mathrm{NA}}  & \mathsmaller{\mathrm{NA}}  & \mathsmaller{\mathrm{NA}}  & \mathsmaller{\mathrm{NA}}  & \mathsmaller{\mathrm{NA}}  \\
\end{bmatrix}
\label{Eq3}
\end{equation}
\noindent where $\mathsmaller{\mathrm{NA}}$ indicates data that cannot be accessed, as they are associated with the  z-polarization, perpendicular to the sample surface. The normalized theoretical coincidence counts obtained with this effective susceptibility tensor are shown in red lines in Fig. \ref{Fig3}a-d, which matches the experimental data. Notably, the susceptibility tensor deviates from the ZB crystalline structure (see Supplementary Information), showing an enhanced relative contribution of $\chi^{(2)}_{xxx}$, associated to the type-0 process under x-axis pumping (H polarization).

\subsection{Quantum state tomography of a single GaAs nanowire}
Although the method to estimate the effective susceptibility tensor does not provide access to absolute values, it enables reliable estimation of the quantum state produced by a single NW. We can indeed predict theoretically the density matrix $\rho_{\mathrm{1}}^{\mathrm{th,P}} =\ket{\Psi_{\mathrm{1}}^{\mathrm{th,P}}}\bra{\Psi_{\mathrm{1}}^{\mathrm{th,P}}} $ for every pump polarization $\mathrm{P}\in $\{H,V,D,A\}, by inserting $\chi_{eff}^{(2)}$ (Eq. \ref{Eq3}) in Eq. \ref{Eq1} and \ref{Eq2}. The theoretical matrices are shown in Figure \ref{Fig3}e-h and confirm a type-0 process when the NW is pumped along its long axis (Fig. \ref{Fig3}e). They also highlight that a single NW emits photons through a type-II process and generates a Bell state $\ket{\Psi^{+}} = \dfrac{1}{\sqrt{2}}(\ket{HV}+\ket{VH})$ if pumped vertically (Fig. \ref{Fig3}f). However, this process has low efficiency, with a coincidence count rate being negligible compared to the H pump, with $\chi^{(2)}_{yij}\ll\chi^{(2)}_{xxx}$ (see Eq. \ref{Eq3}). The coincidence counts rate is indeed an order of magnitude higher for the H pump compared to the V pump, which is confirmed by the theoretical density matrices for the polarization of the D and A pump (Fig. \ref{Fig3}g,h) where the photon pairs are mainly polarized along the long axis of the NW. An efficient generation of a Bell state from type-II SPDC has recently been shown in a similar material composed of a thin film of AlGaAs \cite{Stich25}. However, this process arises from the inherent nonlinear characteristics of the tensor and consequently limits its flexibility. This limitation can be overcome by using two NWs, as will be discussed later.\\ 

\begin{figure}[h!]
   \centering
    \includegraphics[width = \textwidth]{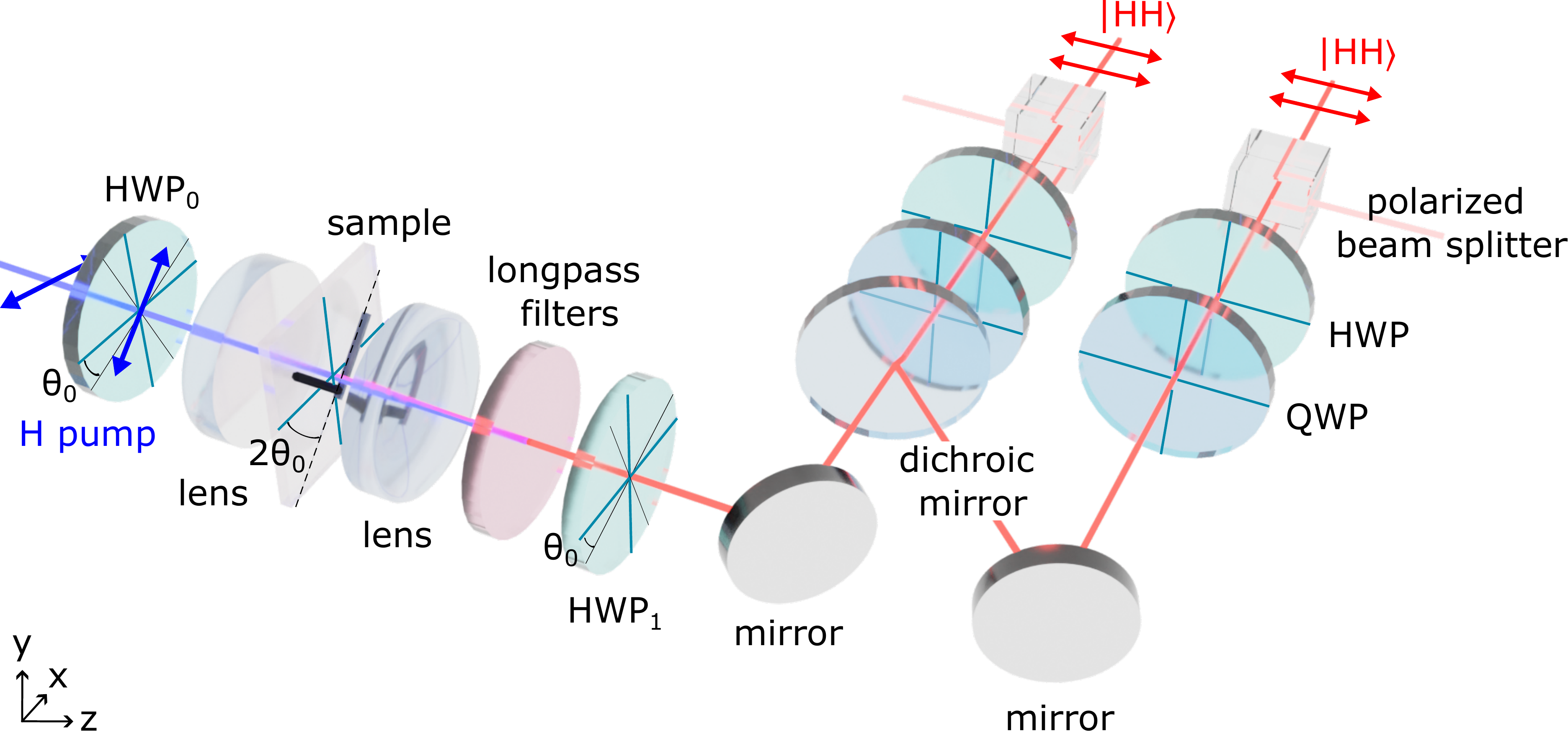}
  \caption{Schematic illustration of the quantum state tomography setup. A lens (f = 8 mm) focuses a CW pump (at 778 nm) on the nanowire(s). The SPDC generated photons are collected with an identical lens, filtered by longpass filters. The signal and idler photons are separated by a dichroic mirror and quarter wave plates (QWP), half wave plates (HWP) and polarized beam splitters (PBSs) are inserted in the two paths. The photons are then coupled to fibers and detected by SNSPDs. The half wave plate $ \mathrm{HWP_{0}}$ controls the pump polarization, $ \mathrm{HWP_{1}}$ projects the photons emitted along the long axis of the nanowire on the horizontal axis. (More details in the Supplementary Information.)}
  \label{Fig4}
\end{figure}

The quantum state of the generated photon pairs can then be measured experimentally by performing quantum state tomography with the setup shown in Fig. \ref{Fig4}. The signal and idler photons are separated by a dichroic mirror with a cutoff wavelength at 1555 nm and two sets of HWPs, quarter wave plates (QWPs) and PBSs are inserted in the signal and idler arms. The addition of these optical components enables projecting the light onto different polarization states and estimating the density matrix of the two-photon state via the maximum likelihood estimation method (Fig.\ref{Fig3}i-l) \cite{james_measurement_2001}. For each projection the signal is integrated for 15 min. As shown in Fig. \ref{Fig3}, the experimental density matrices $\rho_{\mathrm{1}}^{\mathrm{exp,P}}$ agree very well with the theoretical ones derived from the fit, thus confirming the validity of the susceptibility tensor estimation. The experimental imaginary parts of the density matrices as well as the density matrices estimated from the theoretical susceptibility tensor of a bulk ZB crystal 
are provided in the Supplementary Information for conciseness. 

\subsection{Tunable polarization entanglement from two orthogonal nanowires}

Once the single GaAs NW is fully characterized, the quantum state produced by two identical orthogonal NWs (shown in Fig. \ref{Fig1}a,b) can be theoretically predicted by taking into account the contributions of the two NWs with

\begin{equation}
      \ket{\Psi_{\mathrm{\mathrm{2}}}^{\mathrm{th, P}}}(\mathbf{u}) = \ket{\Psi_{\mathrm{\mathrm{NW_{H}}}}^{\mathrm{th, P}}}(\mathbf{u})+\ket{\Psi_{\mathrm{NW_{V}}}^{\mathrm{th, P}}}(\mathbf{u})
      \label{Eq4}
\end{equation}

\noindent where $\ket{\Psi_{\mathrm{\mathrm{NW_{H}}}}^{\mathrm{th, P}}}(\mathbf{u}) =  \ket{\Psi_{\mathrm{1} }^{\mathrm{th, P}}}(\mathbf{u})$ (see Eq. \ref{Eq1}) and $\ket{\Psi_{\mathrm{\mathrm{NW_{V}}}}^{\mathrm{th, P}}}(\mathbf{u})$ is obtained by taking into account the 90$^{\circ}$ rotation of the second NW relative to the first one. Details on the computation can be found in the Methods section. \\

\begin{figure}[h!]
   \centering
    \includegraphics[width = \textwidth]{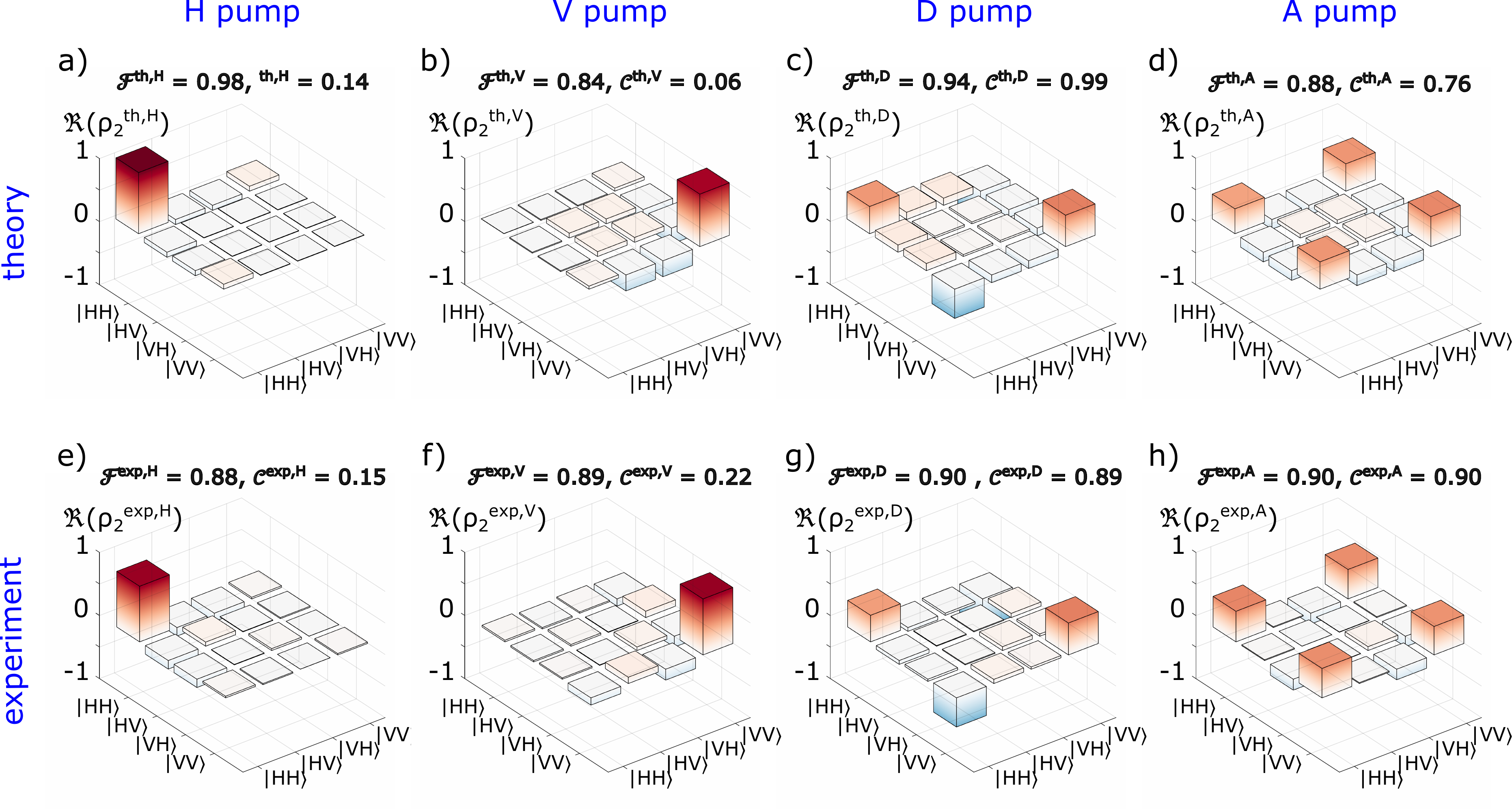}
  \caption{Polarization-entangled Bell state generation from the GaAs nanowires. a-d)  Real part of the theoretical density matrix $\rho_{\mathrm{2}}^{\mathrm{th,P}} =\ket{\Psi_{\mathrm{2}}^{\mathrm{th,P}}}\bra{\Psi_{\mathrm{2}}^{\mathrm{th,P}}} $ obtained with the fitted $\chi^{(2)}$ (Eq. \ref{Eq3}), with P the pump polarization. e-h) Real part of the experimental density matrix $\rho_{\mathrm{2}}^{\mathrm{exp,P}} =\ket{\Psi_{\mathrm{2}}^{\mathrm{th,P}}}\bra{\Psi_{\mathrm{2}}^{\mathrm{th,P}}} $, with P the pump polarization. Each column correspond to a given P pump polarization being horizontal (H) for a),e); vertical (V) for b),f); diagonal (D) for c),g); and antidiagonal (A) for d),h). The fidelities $\mathcal{F}$ and concurrences $\mathcal{C}$ are computed for the comparison with ideal density matrices where the NW would emit pure type-0 SPDC.} 
  \label{Fig5}
\end{figure}

By inserting the susceptibility tensor values in Eqs. \ref{Eq2} and \ref{Eq4}, the theoretical density matrices $\rho_{\mathrm{2}}^{\mathrm{th, P}} =\ket{\Psi_{\mathrm{2}}^{\mathrm{th, P}}}\bra{\Psi_{\mathrm{2}}^{\mathrm{th, P}}} $ can be computed as a function of the pump polarizations $P$, with H the polarization aligned along the long axis of $\mathrm{NW_{H}}$. The real part of the density matrices are plotted in Fig. \ref{Fig5}a-d. To better understand the density matrices, we can consider to a first approximation an ideal type-0 process where only $\chi_{xxx}^{(2)}$ is non zero. In that case, the two photon wave functions would be written as $\ket{\Psi_{2}^{\mathrm{id}}}(\theta) \propto \cos{\big(2(\theta-\theta_{0})\big)}\ket{HH}-\sin{\big(2(\theta-\theta_{0})\big)}\ket{VV}$ with $\theta$ the angle of rotation of $\mathrm{HWP_{0}}$ that sets the pump polarization. The negative sign is chosen for the consistency with the experiment (see Methods). When $\theta=\theta_{0}$ the pump polarization is aligned with the long axis of $\mathrm{NW_{H}}$ (Fig. \ref{Fig1}c). In that case, the photon pair originates mainly from this NW only and $\ket{\Psi_{2}^{\mathrm{id}}}(\theta_{0}) \propto \ket{HH}$. Similarly, when the pump polarization is rotated by 90 $^{\circ}$, the photon pair is generated by the second NW only and $\ket{\Psi_{2}^{\mathrm{id}}}(\theta_{0}+45^{\circ}) \propto \ket{VV}$. This phenomenon can be observed in the theoretical estimation of the density matrices for the H and V pump in Fig. \ref{Fig5}a,b. On the other hand, when the pump is aligned at $\pm 45^{\circ}$ of the NW long axis, then two Bell states are generated with $\ket{\Psi_{2}^{\mathrm{id}}}(\theta_{0}\pm 22.5^{\circ}) \propto \frac{1}{\sqrt{2}}\big(\ket{HH}\mp\ket{VV})$, due to the superposition of the photon pairs emitted by the two NWs, as shown in Fig. \ref{Fig5}c,d. In order to quantify how close the theoretical density matrices $\rho_{\mathrm{2}}^{\mathrm{th, P}} =\ket{\Psi_{\mathrm{2}}^{\mathrm{th, P}}}\bra{\Psi_{\mathrm{2}}^{\mathrm{th, P}}} $ derived from $\chi^{(2)}_{eff}$ are to the ideal cases  $\rho_{\mathrm{2}}^{{\mathrm{id}}} =\ket{\Psi_{2}^{\mathrm{id}}}\bra{\Psi_{2}^{\mathrm{id}}} $, we can compute the fidelities $\mathcal{F(\rho_{\mathrm{2}}^{{\mathrm{id}}},\rho_{\mathrm{2}}^{\mathrm{th, P}})}$ (see Methods) \cite{Jozsa94}. The figures of the ideal type-0 density matrices are shown in the Supplementary Information. As a result, the fidelities are higher than $84\%$ for every pump polarization, confirming that the main contribution is a type-0 process arising from the longitudinal component of the NWs. Similarly, the degree of entanglement can be quantified by computing the concurrence $\mathcal{C}$ as a figure of merit \cite{Wootters98}, ranging from 0 (fully separable) to 1 (fully entangled). The system is shown to be theoretically in a separable state for H and V pump polarizations with $\mathcal{C}^{\mathrm{th,H}}=0.14$ and $\mathcal{C}^{\mathrm{th,V}}=0.06$, and fully entangled for D and A pump polarizations, with $\mathcal{C}^{\mathrm{th,D}}=0.99$ and $\mathcal{C}^{\mathrm{th,A}}=0.76$. The discrepancy between the latter two values can be attributed to some asymetries in the $\chi_{ijk}^{(2)}$ tensor. It should be noted that such tunability cannot be achieved using standard linear polarization components \cite{Lung20}. The orthogonal NWs thus function as a switchable nanoscale source, acting as a universal building block capable of generating entangled or separable states on demand, without the need for additional optical components. This increased degree of freedom could benefit quantum applications such as quantum teleportation \cite{Modlawska08}. In particular, when the qubit undergoes multiple successive teleportations, the resulting quantum errors can be corrected more efficiently when using nonmaximally entangled states.

In order to experimentally perform quantum tomography for the two orthogonal NWs, the same experimental setup was used as for the single NW (Fig. \ref{Fig4}). The horizontal pump polarization, aligned with the first NW ($\mathrm{NW_{H}}$ in Fig. \ref{Fig1}c) has been obtained by maximizing the SHG with $\theta^{H} = \theta_{0}$. Details on the procedure, similar as for the single NW, can be found in the Methods section and in the Supplementary Information. We set $\theta^{V} = (\theta_{0} + 45^{\circ}$) for the vertical pump, to probe the vertical NW ($\mathrm{NW_{V}}$ in Fig. \ref{Fig1}c). However, the intensity of SHG is lower for $\mathrm{NW_{V}}$, a behavior likely arising from its slightly reduced volume. Thus,  the diagonal and antidiagonal pump polarizations were adjusted when performing the SPDC measurements to have the same coincidence rates from the two NWs as well as the highest fidelities. It has been achieved by rotating the pump polarization and adjusting the sample position in order to have equal coincidence counts in the \{HH\} and \{VV\} basis. This has been reached for $\theta^{D} = \theta_{0} + 25^{\circ}$ and $\theta^{A} = \theta_{0} - 23^{\circ}$. The real part of the density matrices $\rho_{\mathrm{2}}^{\mathrm{exp, P}} = \ket{\Psi_{\mathrm{2}}^{\mathrm{exp, P}}}\bra{\Psi_{\mathrm{2}}^{\mathrm{exp, P}}}$, retrieved by maximum likelihood estimation, is plotted in Fig. \ref{Fig5}e-h) as a function of the pump polarization P, again showing a very good agreement between experiment and theory with two separable states for the pump polarization of H and V (Fig. \ref{Fig5}e,f) and two Bell states $\ket{\Psi_{\mathrm{\mathrm{2}}}^{\mathrm{exp, D/A}}} = \frac{1}{\sqrt{2}}\big(\ket{HH}\mp\ket{VV})$ for the pump polarization of D and A (Fig. \ref{Fig5}g,h). The fidelities comparing the density matrices with the ideal type-0 cases are higher than $88\%$, even reaching $(90\pm4)\%$ for the D pump (Fig. \ref{Fig5}g) and $(90\pm4)\%$ for the A pump (Fig. \ref{Fig5}h). The fidelities with the pump along H or V can also reach $95\%$ if the position of the sample is set to maximize the coincidence counts only in the basis \{HH\} or \{VV\}, respectively. The experimental results also confirm that changing the polarization of the pump enables tuning the degree of polarization entanglement, with a concurrence $\mathcal{C}^{\mathrm{exp,H}}=0.15\pm0.07$ and $\mathcal{C}^{\mathrm{exp,V}}=0.22\pm0.09$ for H and V pump, respectively (i.e. separable state) to $\mathcal{C}^{\mathrm{exp,D}}=0.90\pm0.07$ and $\mathcal{C}^{\mathrm{exp,A}}=0.89\pm0.05$ for D and A pump, respectively (i.e. almost maximally entangled photon pairs). The values of the concurrences and the fidelities are reported in Table \ref{Tab}. The slight discrepancy between the theoretical and experimental results can be explained firstly by the fact that the orthogonal NWs are not perfectly identical and are not the same as the one used to obtain the effective susceptibility tensor. Secondly, the vertical NW is not perfectly aligned at $90^{\circ}$ with respect to the first one (see Method), a deviation that is not taken into account in the theoretical estimation. Finally, the theoretical method does not provide information on the imaginary part of the density matrix since the effective susceptibility tensor is real per construction. Even though the imaginary experimental parts of the density matrices are small, this results in a minor difference. 

\begin{table}[h]      
\centering    
\begin{tabular}{c|c|c|c|c|c}
\multicolumn{1}{c}{} & & Pump H & Pump V & Pump D & Pump A \\
\bottomrule
\multirow{2}{*}{Fidelity (\%)} & $\mathcal{F^{\mathrm{th}}}$ & 98 & 84 & 94 & 88 \\
& $\mathcal{F^{\mathrm{exp}}}$ &88 & 89 & 90 & 90 \\
\cmidrule(lr){1-6} 
\multirow{2}{*}{Concurrence} & $\mathcal{C^{\mathrm{th}}}$ & 0.14 & 0.06 & 0.99 & 0.76 \\
& $\mathcal{C^{\mathrm{exp}}}$ & 0.15 & 0.22 & 0.89 & 0.90 \\
\bottomrule
\end{tabular}
\caption{Fidelities and concurrences evaluated when comparing the theoretical and experimental density matrices shown in Fig. \ref{Fig5} of the orthogonal nanowires (\ref{Fig1}c) with the ideal type-0 density matrices; as a function of the pump polarization P.} 
\vspace{-3em} 
\label{Tab}
\end{table}

We have shown that an independent fit on a single NW enables us to extract the effective susceptibility tensor of GaAs NWs and to reconstruct the density matrices of the quantum states produced by SPDC for both a single NW and two orthogonal NWs, as a function of the pump polarization. 

\section{Conclusion}\label{sec3}
In this work, we have exploited type-0 SPDC in orthogonally oriented Zinc Blende GaAs nanowires, pumped simultaneously, to generate polarization entangled states at the nanoscale. This work provides both a tunable quantum source of entangled state at telecommunication wavelengths and a robust approach for determining the relative susceptibility tensor values from coincidence count fits of photon pairs emitted by a single NW. This enabled us to infer the theoretical density matrices of the photon pairs generated by two orthogonal nanowires, showing excellent agreement with the experimental data. This nanosource also offers greater flexibility: tuning the pump polarization indeed enables us to tune the quantum state from separable to fully entangled, without additional optical components. This increased degree of freedom has the potential to support advanced quantum functionalities, such as quantum teleportation. 
Leveraging bottom-up fabrication as well as nanowire movability could improve the efficiency and performance of the quantum source of entangled photon pairs. More sophisticated architectures made of NWs could indeed be fabricated, either grown on a photonic chip or accurately placed on a chip or inside a cavity using a micromanipulator or an atomic force microscopy tip. A spatial light modulator (SLM) could also be used to adjust the pump intensity on each nanowire. Furthermore, an assembly of orthogonal Langmuir-Blodgett NWs \cite{Tabassom22} could be explored to enhance the brightness of the quantum source. 

\section{Acknowledgment}
The authors thank the Scientific Center for Optical and Electron Microscopy (ScopeM), the Binning and Rohrer Nanotechnology Center (BRNC), the FIRST cleanroom and D-MATL X-ray service platform at ETH Zurich for technical assistance. This work was supported by the Swiss National Science Foundation SNSF (Consolidator Grant 213713 and Grant 179099). the authors also thank the Nanolyon platform, a member of the CNRS-RENATECH+ French national network, for access to equipment and J. B. Goure for technical assistance. We also thank Eleni Prountzou and Ülle-Linda Talts for their technical assistance with the SEM images.\\
The authors also acknowledge Ryo Mizuta Graphics for providing the resources used for 3D images. This asset is used with permission.

\section{Author contributions statement}

E.B-R., R.G., G.S. and R.J.C. designed the study. G.S., A. M., W.J., E.B-R. and Z.P. built the setup. E.B-R. and Z.P. conducted the experiments. E.B-R performed the numerical simulations. E.B-R, Z.P.,  R.J.C and R.G. analysed the data. T.D., P.R., A.D., N.C., and J.P. fabricated the GaAs NWs. G.S. transferred mechanically the NWs on a glass substrate. E.B-R. and H.W. used a micromanipulator to remove neighbouring nanowires. R.G. supervised the research. E.B-R wrote the original manuscript. All authors reviewed the manuscript and gave approval to the final version of the manuscript. 

\section{Methods}

\subsection{Nanowire fabrication}
The GaAs NWs come from the same sample of ref. \cite{saerens_background-free_2023} and have been fabricated by the Lyon Institute of Nanotechnology (INL) team. The NWs were grown by molecular beam epitaxy (MBE) on epi-ready Si(111) subtrates, using the vapor-liquid-solid (VLS) mechanism. Further details on the VLS growth are available in the Supplementary Information of ref. \cite{saerens_background-free_2023}. After growth, the NWs were mechanically transferred onto a quartz substrate coated with a 10 nm thick ITO layer (to enable SEM imaging), where they lay flat on the substrate.\\ 
The single reference NW (Fig. \ref{Fig1}b) is one of the NWs previously investigated in ref. \cite{saerens_background-free_2023}. It has a tapered shape, with a length of $4.5\pm0.1$ \textmu m and a diameter of $\approx 400$ nm.\\
The sample has then been scanned to select nanowires oriented orthogonally (Fig. \ref{Fig1}c). The horizontal ($\mathrm{NW_{H}}$) and vertical ($\mathrm{NW_{V}}$) nanowires are both tapered and exhibit lengths of $4.7\pm0.1$ \textmu m  and $4.5\pm0.1$ \textmu m respectively, with diameters of $\approx$ 400 nm. Subsequently, a micromanipulator was used to remove neighboring nanowires or dust particles, thereby minimizing accidental counts and background noise. 

\subsection{Photon pair generation measurements}
A continuous-wave (CW) laser operating at a wavelength of 778 nm (Toptica DL pro 780) was used as a pump source, operating at a power of 12 mW. The beam was focused onto the sample using a lens (Thorlabs A240TM, f = 8 mm, NA = 0.5) producing a spot with an FWHM of approximately 4 \textmu m. The polarization of the pump was controlled with a half wave plate (HWP) (Thorlabs WPH05M-780), and the generated photon pairs were collected by an identical lens. The pump light and possible fluorescence signal were cut by three long pass filters (Semrock LP1064, Semrock LP1319, and Thorlabs FELH1500) placed in the collection path. The complete schematics of the two setups are shown in the Supplementary Information.

\subsubsection{Characterization of a single NW}
Different optical elements were then inserted in the detection path to perform projection measurements for the estimation of the susceptibility tensor experiment. An additional HWP ($\mathrm{HWP_{2}}$) was placed in the detection path to rotate the polarization of the quantum state. Since the photons are coupled to fibers that do not preserve polarization and the SNSPDs are sensitive to polarization, a polarized beam splitter (PBS) was inserted before the fibers to project the quantum state onto the horizontal axis (see Fig. \ref{Fig2}). The photons were then coupled into a single mode fiber (SMF-28) and  separated by a 50:50 fiber splitter, before being detected by two independent superconducting nanowire single-photon detectors. The coincidence counts (corrected from the background) were then measured using a time-to-digital converter. 

\subsubsection{Quantum state tomography}
For the quantum state tomography measurements, the fiber splitter of the previous setup was replaced by a tilted broadband short pass filter (Edmund Optics SP1600) serving as a dichroic mirror, with a cutoff wavelength of 1555 nm to deterministically separate the signal and idler photons. A quarter wave plate, a HWP and a PBS were then inserted in each arm after the dichroic mirror, prior to fiber coupling, to perform the projection measurements. Then, the photons were detected by two SNSPDs. \\
Because the SNSPDs exhibit polarization dependent detection efficiency, the setup was first calibrated with an attenuated 1555 nm laser. Two 3-Paddle polarization controllers were used to maximize the detection efficiencies of the SNSPDs, and the coincidence counts were measured via a time-to-digital converter.\\\
The number of coincidence counts (corrected from
background) provides a dataset of 16 measurements that enables to estimate the density matrix of our two-photon state via the maximum likelihood estimation method. The binwidth is chosen to be 50 ps, the number of bins is 400, and the coincidence window has a width of 550 ps. The density matrix is retrieved using the publicly available tomography library developed by the Kwiat group (https://research.physics.illinois.edu/QI/Photonics/tomography/) \cite{james_measurement_2001}. The datasets are shown in the Supplementary Information for a single NW under H pump polarization as well as for the two orthogonal NWs when pumped by a A pump. \\

In order to experimentally perform quantum tomography of the state generated by the two orthogonal NWs, it is first necessary to determine the angle of $\mathrm{HWP_{0}}$ which enables pumping the sample along the long axis of the first NW ($\mathrm{NW_{H}}$ in Fig. \ref{Fig1}c). As for the single NW, this angle has been found when the SHG is maximal for $\mathrm{NW_{H}}$ and this position has been reached for $\theta^{H} = \theta_{0}$. For this specific pump polarization, the SHG contribution of the second NW is negligible (see Supplementary Information). This angle then serves as a reference for pumping the two NWs with horizontal polarization when performing SPDC measurement. Similarly to the single NW case, the axis of $\mathrm{HWP_{1}}$ has been rotated by the same angle $\theta_{0}$ to project the long axis of the $\mathrm{NW_{H}}$ on the horizontal x-axis (See Fig. \ref{Fig4}). By using the same SHG analysis on the second NW ($\mathrm{NW_{V}}$ in Fig. \ref{Fig1}c), the intensity has been maximized for $\theta^{V} = (\theta_{0} + 44^{\circ})\pm 2^{\circ}$, confirming that the 2 NWs are nearly orthogonal.\\

The comparison between the theoretical density matrix $\rho_{\mathrm{2}}^{\mathrm{th, P}}$ (computed from the fitted susceptibility tensor) and the ideal density matrix $\rho_{\mathrm{2}}^{{\mathrm{id}}}$ (computed for an ideal type-0 process) is achieved by computing the fidelities, defined as 
\begin{equation}
\mathcal{F(\rho_{\mathrm{2}}^{{\mathrm{id}}},\rho_{\mathrm{2}}^{\mathrm{th, P}})}=\left(\mathrm{Tr}\big(\sqrt{\sqrt{\rho_{\mathrm{2}}^{\mathrm{th, P}}}\rho_{\mathrm{2}}^{\mathrm{id}}\sqrt{\rho_{\mathrm{2}}^{\mathrm{th, P}}}}\big)\right)^2.
\end{equation}

\subsection{SHG measurements}
A pulsed laser (Menhir Photonics, MENHIR-1550) was used to generate femtosecond pulses with a central wavelength of 1555 nm, a pulse duration of $\approx 150$ fs and a repetition rate of 217 MHz. The laser is focused onto the sample using a lens (Thorlabs A240TM, f= 8 mm, NA= 0.5). The resulting SHG signal was collected by an identical lens and detected with a visible-range CMOS camera (Andor Zyla sCMOS). To block the pump, two short pass filters were used (Thorlabs FESH0900). The average pump power was set to 9mW and its polarization was controlled using a HWP before the sample.

\subsection{Cost function and fit of the susceptibility tensor}

In order to extract the effective susceptibility tensor values of a single GaAs NW, projection measurements have been performed on the quantum state, by inserting a HWP (labelled $\mathrm{HWP_{2}}$ in Fig. \ref{Fig2}) and a PBS between the sample and the detectors, as shown in Fig. \ref{Fig2}. The experimental coincidence counts $N_{\mathrm{cc}}^{\mathrm{exp,P}}$ have been measured as a function of the rotation angle $\alpha$ of $\mathrm{HWP_{2}}$, and as a function of the pump polarization P, set by $\mathrm{HWP_{0}}$ before the sample (see Fig. \ref{Fig2}). $\mathrm{HWP_{0}}$ was varied among four distinct orientations so that it would be either aligned with the long axis of the NW (referred to as the horizontal (H) pump polarization), perpendicular to the long axis (referred to as the vertical (V) pump polarization) or diagonal and antidiagonal to the long axis (referred to as diagonal (D) and antidiagonal (A) pump polarizations).\\

The theoretical coincidence counts $N_{\mathrm{cc}}^{\mathrm{th, P}}$ can be predicted by taking into account the optical elements in the detection path as well as the quantum state $\ket{\Psi_{\mathrm{1} }^{\mathrm{th, P}}}$ generated from a single NW by SPDC, as a function of the P pump polarization ($\mathrm{P}\in$\{H, V, D, A\}) :

\begin{equation}
N_{\mathrm{cc}}^{\mathrm{th,P}}(\alpha,\mathbf{u}) \propto \Big|\bra{HH}(J_{\mathrm{HWP}}(\alpha)\otimes J_{\mathrm{HWP}}(\alpha)) (J_{\mathrm{M}}\otimes J_{\mathrm{M}})\ket{\Psi_{\mathrm{1} }^{\mathrm{th, P}}}(\mathbf{u})\Big|^2,
\end{equation}

with $J_{\mathrm{HWP}}(\alpha)$ the Jones matrices of the half wave plate

\begin{equation}
J_{\mathrm{HWP}}(\alpha) = 
\begin{bmatrix}
\cos(2\alpha) & \sin(2\alpha) \\
\sin(2\alpha) &-\cos(2\alpha)
\end{bmatrix}
\end{equation}

and $J_{\mathrm{M}}$ the Jones matrix of the mirror 
\begin{equation}
J_{\mathrm{M}} = 
\begin{bmatrix}
-r_{p} & 0 \\
0 & r_{s}  \\
\end{bmatrix}
\end{equation}

where $r_{p}$ and $r_{s}$ are the reflection coefficients for transverse magnetic (TM) and transverse electric (TE) polarizations respectively. For a silver mirror tilted at 45$^{\circ}$ the estimated values are $r_{p} = -0.96 - 0.25i$ and $  r_{s} = -0.99- 0.13i$.\\

We define the SPDC-generated quantum state generated by a single NW by 
\begin{equation}
     \ket{\Psi_{\mathrm{1} }^{\mathrm{th, P}}}(\mathbf{u}) = \sum_{i\in \{x,y\}}E_{i}^{\mathrm{P}}\mathbf{u}_{i},
     \label{Eq9}
\end{equation}
where $\mathbf{u} =[\mathbf{u}_{x} ~ \mathbf{u}_{y}]$, is related to the susceptibility tensor via $\mathbf{u}_{i} =
     \begin{bmatrix}
u_{i1} \\
u_{i2} \\
u_{i3} \\
u_{i4} 
\end{bmatrix} \propto
     \begin{bmatrix}
\chi^{(2)}_{\mathrm{ixx}} \\
\chi^{(2)}_{\mathrm{ixy}} \\
\chi^{(2)}_{\mathrm{iyx}} \\
\chi^{(2)}_{\mathrm{iyy}}
\end{bmatrix}$ in the \{HH, HV, VH, VV\} basis, with $\chi^{(2)}_{\mathrm{ijk}}$ the nonlinear susceptibility tensor values of the GaAs NW. $E_{i}^{\mathrm{P}}$ corresponds to the normalized pump field amplitude with
\[
\begin{cases}
    (E_{x}^{\mathrm{H}},E_{y}^{\mathrm{H}}) = (1,0) \mathrm{~ for~ H~  polarization},\\
    (E_{x}^{\mathrm{V}},E_{y}^{\mathrm{V}}) = (0,1) \mathrm{~ for~ V~  polarization},\\
    (E_{x}^{\mathrm{D}},E_{y}^{\mathrm{D}}) = \dfrac{1}{\sqrt{2}}(1,1) \mathrm{~ for~ D~ polarization}, \\
    (E_{x}^{\mathrm{A}},E_{y}^{\mathrm{A}}) = \dfrac{1}{\sqrt{2}}(1,-1) \mathrm{~ for~ A~ polarization}.  
\end{cases}
\]
We impose the equality $u_{i2} = u_{i3}$ to reflect the indistinguishability of the two photons pairs $\ket{HV}$ and $\ket{VH}$. We also impose $u_{x1} =1$ to ease the relative comparison of the tensor values. \\

To estimate the relative tensor values, we use the experimental coincidence counts $N_{\mathrm{cc}}^{\mathrm{exp,P}}$ presented in Fig. \ref{Fig3}a-d) as a function of the pump polarization $P$ and we define a cost function $f_{\mathrm{cost}}$ that evaluate the difference between the experimental and theoretical coincidence counts, normalized by the corresponding values at a chosen reference angle $\alpha_{\mathrm{max}}$ :

\begin{equation}
f_{\mathrm{cost}}(\mathbf{u}) = \sum_{\alpha,P}\Big|\dfrac{N_{\mathrm{cc}}^{\mathrm{exp,P}}(\alpha)}{N_{\mathrm{cc}}^{\mathrm{exp,P}}(\alpha_{\mathrm{max}})}-\dfrac{N_{\mathrm{cc}}^{\mathrm{th,P}}(\alpha,\mathbf{u})}{N_{\mathrm{cc}}^{\mathrm{th,P}}(\alpha_{\mathrm{max}},\mathbf{u})}\Big|^2.
\label{Eq10}
\end{equation}


Minimizing the cost function with respect to $\mathbf{u}$ yields the values of $\mathbf{u}$ corresponding to the effective susceptibility tensor values given in Eq.\ref{Eq3}. The D and A pump polarizations impose constraints on the minimization problem, with $\ket{\Psi_{\mathrm{1} }^{\mathrm{th, D/A}}}(\mathbf{u}) = \dfrac{1}{\sqrt{2}}(\mathbf{u}_{x}\pm \mathbf{u}_{y})$.

This method does not provide access to the absolute values of the susceptibility tensor, which, however, are not needed to estimate the quantum state and reconstruct the density matrix. \\

\subsection{Quantum state of two orthogonal NWs}

Once the susceptibility tensor is known, one can theoretically predict the density matrix associated with the two orthogonal NWs $\rho_{\mathrm{2}}^{\mathrm{th, P}} =\ket{\Psi_{\mathrm{2}}^{\mathrm{th, P}}}\bra{\Psi_{\mathrm{2}}^{\mathrm{th, P}}} $, by incorporating the quantum states of each NW, $\ket{\Psi_{\mathrm{\mathrm{NW_{H}}}}^{\mathrm{th, P}}} (=  \ket{\Psi_{\mathrm{1} }^{\mathrm{th, P}}}$) for the horizontal and $\ket{\Psi_{\mathrm{NW_{V}}}^{\mathrm{th, P}}}$ for the vertical one : 

\begin{align}
     \ket{\Psi_{\mathrm{\mathrm{2}}}^{\mathrm{th, P}}}(\mathbf{u}) &= \ket{\Psi_{\mathrm{\mathrm{NW_{H}}}}^{\mathrm{th, P}}}(\mathbf{u})+\ket{\Psi_{\mathrm{NW_{V}}}^{\mathrm{th, P}}}(\mathbf{u})\\
      & = \sum_{i\in \{x,y\},j =(x+y)-i}E_{i}^{\mathrm{P}}\mathbf{u}_{i} + E_{j}^{\mathrm{P\perp}}M\mathbf{u}_{j}
      \label{Eq12}
\end{align}

\noindent where $E_{i}^{\mathrm{P}}$ and $\mathbf{u}$ have been previously defined in Eq. \ref{Eq9}, and $E_{j}^{\mathrm{P\perp}}$ corresponds to the orthogonal pump field amplitude with 
\[
\begin{cases}
    (E_{x}^{\mathrm{H}\perp},E_{y}^{\mathrm{H}\perp}) = (0,1),\\
    (E_{x}^{\mathrm{V}\perp},E_{y}^{\mathrm{V}\perp}) = (-1,0),\\
    (E_{x}^{\mathrm{D}\perp},E_{y}^{\mathrm{D}\perp}) = \dfrac{1}{\sqrt{2}}(-1,1), \\
    (E_{x}^{\mathrm{A}\perp},E_{y}^{\mathrm{A}\perp}) = \dfrac{1}{\sqrt{2}}(1,1).  
\end{cases}
\]
and M corresponds to the permutation matrix 
\begin{equation} M =
\begin{bmatrix}
0 &0 &0 & 1\\
0 &0 &1 &0\\
0 & 1 & 0 & 0\\
1 & 0 & 0 & 0
\end{bmatrix}.
\end{equation}

It should be noted that the sign of $\ket{\Psi_{\mathrm{NW_{V}}}^{\mathrm{th, P}}}$ in Eq. \ref{Eq12} is fixed by a change of the basis of the vertical NW to take into account the growth direction, which is consistent with the SEM image of Fig.\ref{Fig1}c. Choosing the opposite sign would have exchanged the diagonal and antidiagonal matrices. 

\newpage
\begin{appendices}

\section{Tunable Near-Infrared Polarization-Entangled
Photons From Orthogonal GaAs Nanowires}\label{secA1}
\subsection{Details on the Experimental setups}
This section gives details on the experimental setups used for characterization of the NWs. 

\begin{figure}[h!]
   \centering
    \includegraphics[width = \textwidth]{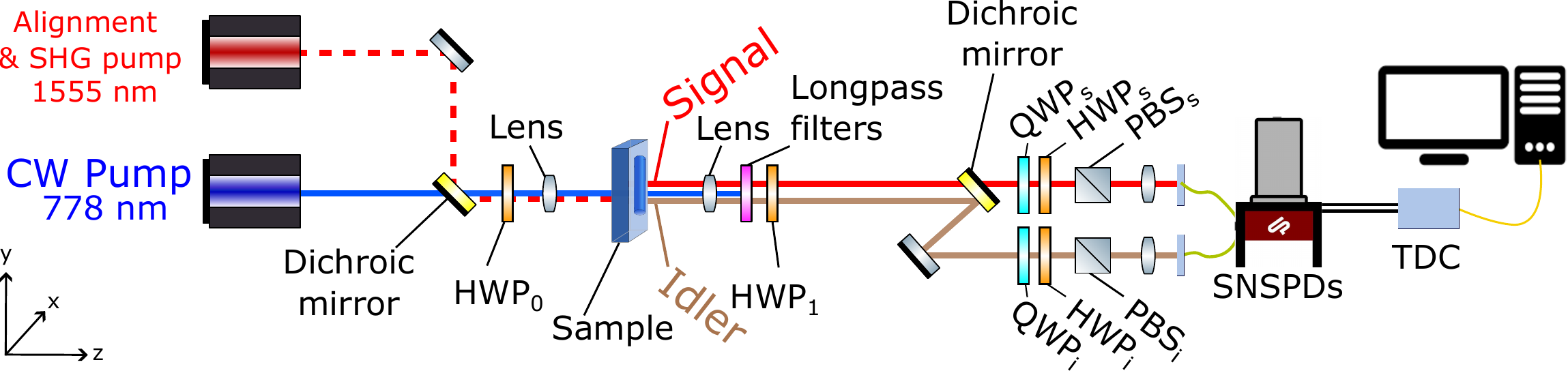}
  \caption{Characterization of a single NW with a Hanburry Brown and Twiss (HBT) setup. A lens (f= 8 mm) focuses a CW pump (at 778 nm) on the nanowire. The SPDC generated photons are collected with a similar lens (f= 8 mm), filtered by three longpass filters. The half wave plate $ \mathrm{HWP_{0}}$ controls the pump polarization, $ \mathrm{HWP_{1}}$ projects the long axis of the nanowire on the horizontal x-axis while $ \mathrm{HWP_{2}}$ rotates the polarization of the quantum state. The polarized beam splitter (PBS) transmits only the horizontal polarization. The idler and signal photons are then coupled to a fiber and detected by superconducting nanowire single-photon detector (SNSPDs).} 
  \label{Fig_S1}
\end{figure}

\begin{itemize}
    \item \underline{Single NW characterization setup}
\end{itemize}
The characterization of the single reference NW has been performed with the setup given in Fig. \ref{Fig_S1}.

\begin{figure}[h!]
   \centering
    \includegraphics[width = \textwidth]{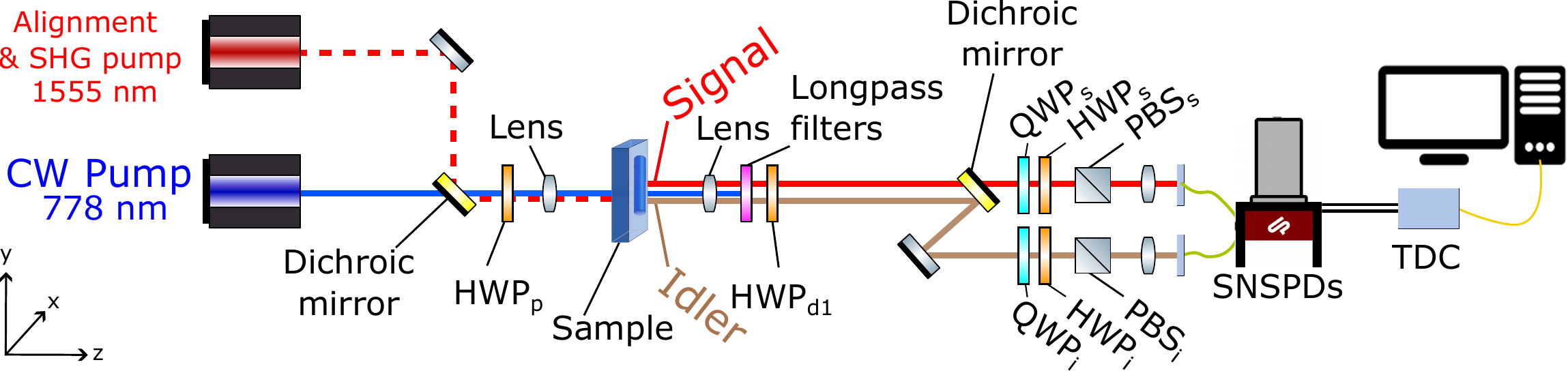}
  \caption{Schematic illustration of the quantum state tomography setup. A lens (f= 8 mm) focuses a CW pump (at 778 nm) on the nanowire(s). The photons pairs generated by SPDC are collected with a similar lens (f= 8 mm), filtered by three longpass filters. The signal and idler photons are separated by a dichroic mirror. Quarter wave plates (QWPs), half wave plates (HWPs) and polarized beam splitters (PBSs) are inserted in the two paths to perform projection measurements. The photons are then coupled to fibers and detected by superconducting nanowire single-photon detector (SNSPDs) and the coincidences are obtained by a time-to-digital converter (TDC). The half wave plate $ \mathrm{HWP_{0}}$ controls the pump polarization, $ \mathrm{HWP_{1}}$ projects the long axis of the nanowire on the horizontal x-axis. The transmission in the fiber and the detection efficiency are optimized using an attenuated 1555 nm laser. } 
  \label{Fig_S2}
\end{figure}

\begin{itemize}
    \item \underline{Quantum state tomography setup}
\end{itemize}
    The density matrices have been retrieved by maximum likelihood estimation from projection measurements obtained with the setup given in Fig. \ref{Fig_S2}.
    
\newpage

\subsection{SHG measurements}
\begin{figure}[h!]
   \centering
    \includegraphics[width = 0.7\textwidth]{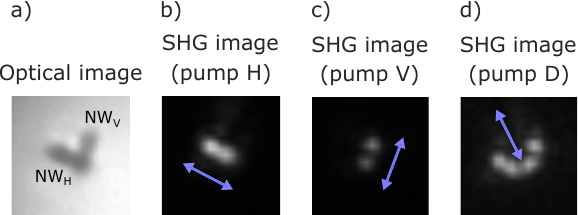}
  \caption{SHG images of orthogonal NWs. a) Optical image of the sample, b) SHG intensity for H pump polarization, c) SHG intensity for V pump polarization and d) SHG intensity for D pump polarization. The pump polarization is represented by the blue arrow. } 
  \label{Fig_SHG}
\end{figure}

    The optical image of the orthogonal NWs as well as the SHG intensities as a function of the pump polarization are shown in Fig. \ref{Fig_SHG}. The pump polarization was rotated according to the angle $\theta$ of $\mathrm{HWP_{0}}$ (see Fig.\ref{Fig2}) to identify the pump orientation aligned with the NWs'long axis. Maximum SHG from  $\mathrm{NW_{H}}$ corresponds to a horizontal pump polarization at $\theta^{H} = \theta_{0}$, for which the contribution of the second NW is negligible, as can be seen in  \ref{Fig_SHG}b). \\
    Applying the same SHG analysis to the second NW ($\mathrm{NW_{V}}$ in Fig. \ref{Fig_SHG}c) yields a maximum at $\theta^{V} = (\theta_{0} + 44^{\circ})\pm 2^{\circ}$, confirming that the 2 NWs are nearly orthogonal. We also notice that the intensity is lower for the V pump, which can be attributed to the slightly lower volume of $\mathrm{NW_{V}}$ compared to $\mathrm{NW_{H}}$.\\
    Fig. \ref{Fig_SHG}d) shows the diagonal pump polarization at which the SHG intensities of both NWs are equal. For the SPDC measurement, this D polarization angle was chosen to equalize the coincidence counts from the two NWs (detected in the bases $\{HH\}$ and $\{VV\}$, respectively).

\subsection{Power sweep}

\begin{figure}[h!]
   \centering
    \includegraphics[width = 0.85\textwidth]{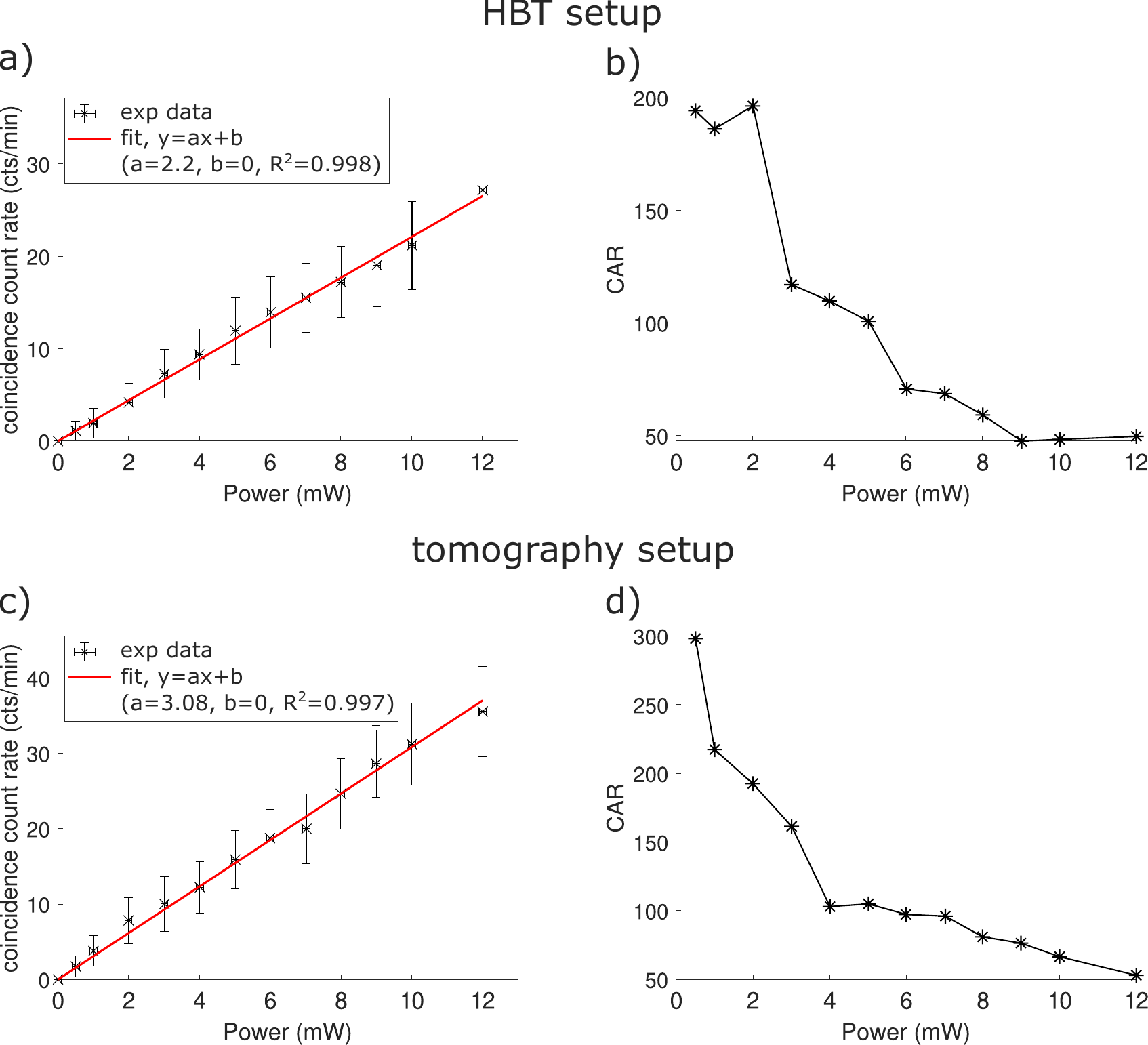}
  \caption{Coincidence count rates (a,c) and corresponding CAR (b,d) estimated for the single reference NW displayed in Fig. \ref{Fig1}b. The fit applied to the coincidence counts (represented by the red line) confirms a linear dependence with the pump power.   The data were acquired with the HBT setup for a,b) and with the tomography setup for c,d).} 
  \label{Fig_S3}
\end{figure}

The SPDC process has been confirmed by the linear increase in photon pairs with respect to the incoming power measured in the HBT setup (see Fig. \ref{Fig_S3}a) and with the tomography setup (see Fig. \ref{Fig_S3}c) for the single reference NW presented in Fig. \ref{Fig1}b when pumped along its long axis (associated with H pump polarization). The measurements were performed in both cases with an integration time of 15 min and a binwidth of 50 ps. The coincidence window at zero delay is 550 ps. \\
The CAR (Coincidence to Accidental ratio) has also been estimated as a function of the incident power, defined as 
\begin{equation}
    \mathrm{CAR} = \frac{R_{\mathrm{tot}-}R_{\mathrm{acc}}}{R_{\mathrm{acc}}},
\end{equation}

where $R_{\mathrm{tot}}$ denotes the total coincidence count rate measured within the coincidence window and $R_{\mathrm{acc}}$ represents the accidental coincidence rate, that is, coincidence events detected within the coincidence window but not originating from the SPDC. The latter has been estimated from the background signal, measured at large time delays between the two detectors, away from the coincidence peak. The CAR is related to the normalized second-order correlation function $g^{(2)}(0)$ with $\mathrm{CAR} = g^{(2)}(0)-1$.\\
The CAR was characterized in both the HBT setup (see Fig. \ref{Fig_S3}b) and the tomography setup (see Fig. \ref{Fig_S3}d) as a function of the pump power. The data follows the expected dependence on pump power, since the true coincidence counts $R_{\mathrm{tot}-}R_{\mathrm{acc}}$ scale linearly with the pump power while the accidental counts $R_{\mathrm{acc}}$ scale quadratically. For pump powers below 2 mW, the CAR exceeds 150 in both measurement conditions. 

\newpage
\subsection{Projection measurements in HBT Setup}
\begin{figure}[h!]
   \centering
    \includegraphics[width = 0.7\textwidth]{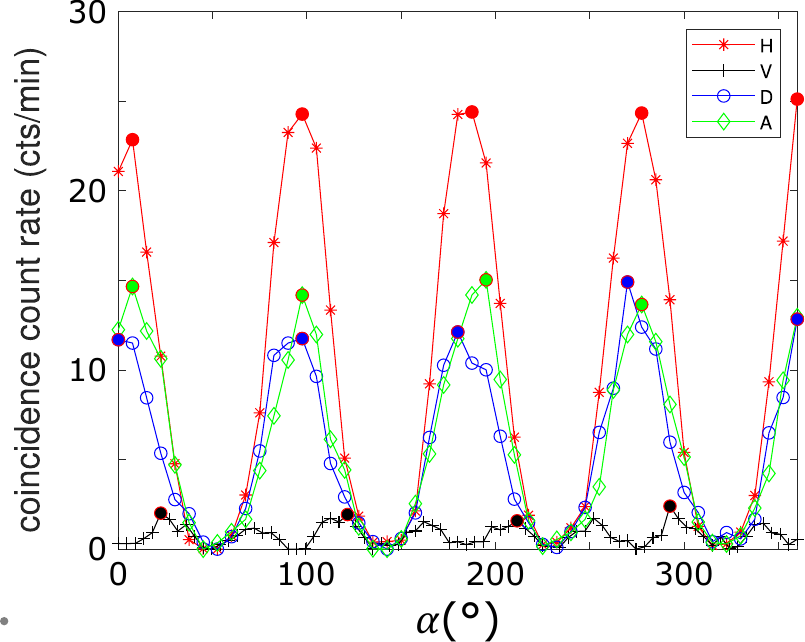}
  \caption{Coincidence count rates for the reference single NW, presented in Fig. \ref{Fig1}b), as a function of the pump polarization. The polarization considered are H (horizontal, whith the pump aligned along the NW's long axis, in red), V (vertical, at 90$^{\circ}$ to H, in black), and D/A diagonal and antidiagonal pump polarizations, in blue and green, respectively). The maxima, indicated by solid dots circled in red, are used to computed the mean values of the maxima, from which the ratio between the H and other polarizations is estimated $\mathcal{R}^{H/P}=\dfrac{\langle\mathrm{max}(N_{\mathrm{cc}}^{\mathrm{exp,H}})\rangle}{\langle\mathrm{max}(N_{\mathrm{cc}}^{\mathrm{exp,P}})\rangle}$, with $\langle.\rangle$ indicating the mean value. The resulting ratio are $\mathcal{R}^{H/V}=12.4$, $\mathcal{R}^{H/D}=1.9$ and $\mathcal{R}^{H/A}=1.7$. }
  \label{Fig_S4}
\end{figure}

Projection measurements were carried out to estimate the effective susceptibility tensor of a single GaAs NW, as detailed in the main manuscript. Coincidence counts were recorded as a function of the rotation angle $\alpha$ of $\mathrm{HWP_{2}}$,  (see Fig. \ref{Fig_S1} and Fig. \ref{Fig2}) as well as of the pump polarization. The unormalized coincidence count rates plotted as a function of $\alpha$ (shown in Fig. \ref{Fig_S4}). The ratio between the H and other polarizations has been estimated using the maxima of the coincidence count rates, indicated by solid dots circled in red in Fig. \ref{Fig_S4}. It is defined as 
\begin{equation}
    \mathcal{R}^{H/P}=\dfrac{\langle\mathrm{max}(N_{\mathrm{cc}}^{\mathrm{exp,H}})\rangle}{\langle\mathrm{max}(N_{\mathrm{cc}}^{\mathrm{exp,P}})\rangle},
\end{equation} with $\langle.\rangle$ indicating the mean value. The resulting ratios are $\mathcal{R}^{H/V}=12.4$, $\mathcal{R}^{H/D}=1.9$ and $\mathcal{R}^{H/A}=1.7$. These results indicate that photon-pair generation with a horizontally polarized pump is more than an order of magnitude more efficient than with a vertically polarized pump. 

\clearpage
\subsection{Quantum state tomography}
\subsubsection{Quantum tomography for the reference single NW}
\begin{figure}[h!]
   \centering
    \includegraphics[width = 1\textwidth]{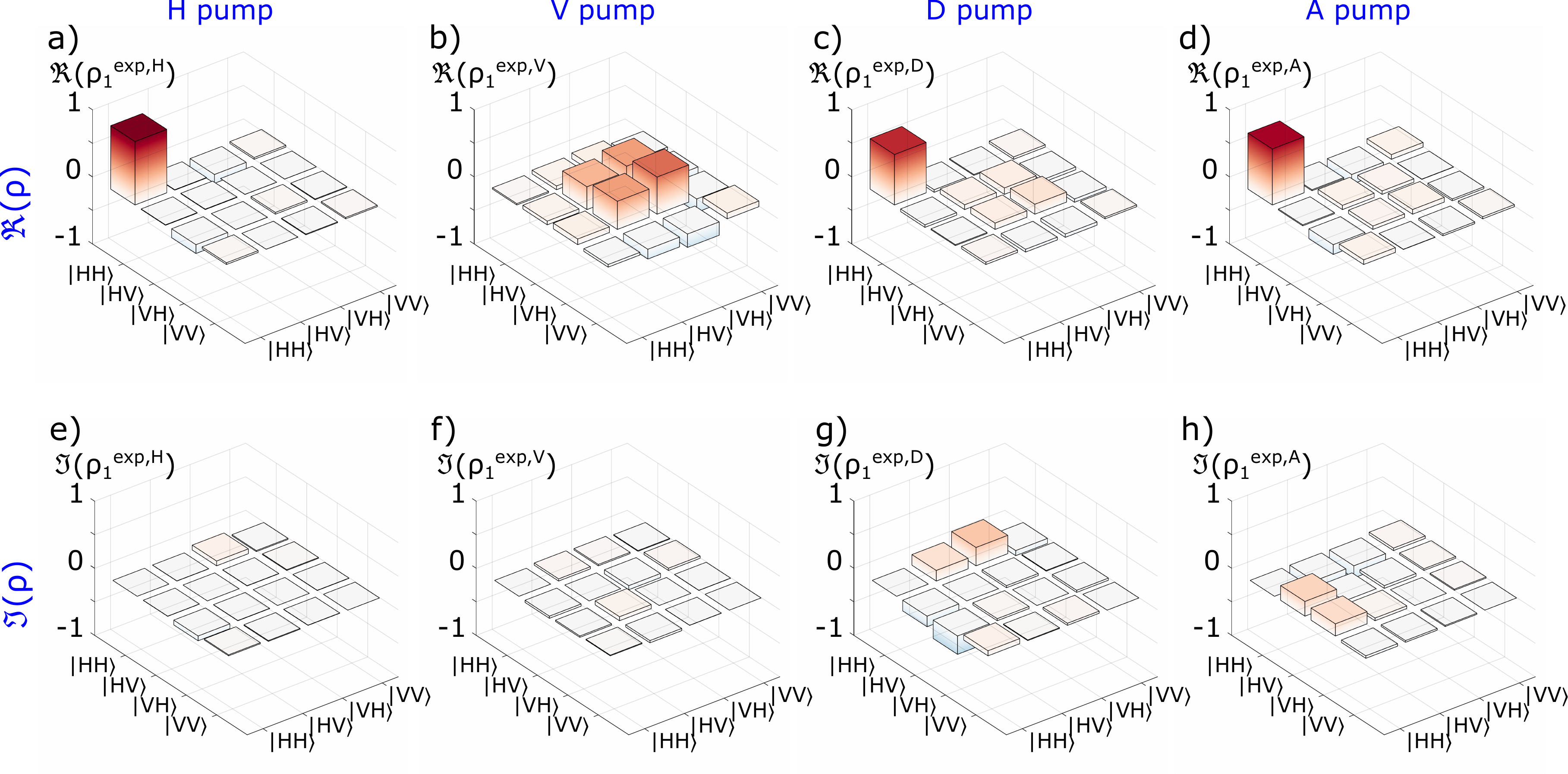}
  \caption{Experimental density matrice $\rho_{\mathrm{1}}^{\mathrm{exp,P}}$ measured for the single reference NW presented in Fig. \ref{Fig1}b. Each column correspond to a given P pump polarization being horizontal (H) for a),e); vertical (V) for b),f); diagonal (D) for c),g); and antidiagonal (A) for d),h). The real parts of the density matrices are shown in the first row (a-d), and the imaginary parts are shown in the second row (e-h).} 
  \label{Fif_S5}
\end{figure}

\begin{figure}[h!]
   \centering
    \includegraphics[width = 1\textwidth]{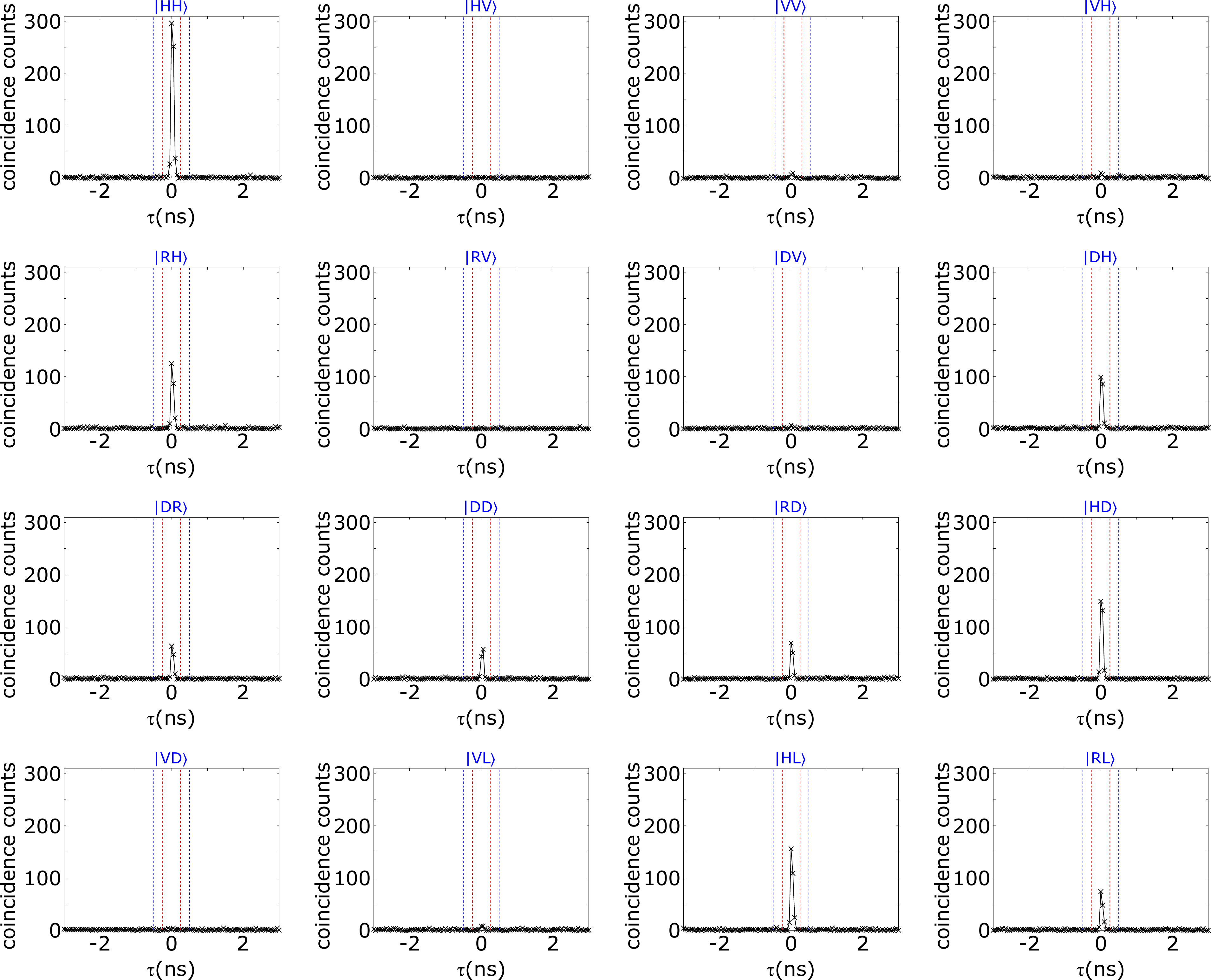}
  \caption{Experimental coincidence histograms of quantum state tomography for the reference single NW, for H pump polarization. The vertical red dashed lines indicate the coincidence window, while the blue dashed lines represent the area to exclude when calculating the background. } 
  \label{Fif_S5b}
\end{figure}

We show in this section the real and imaginary parts of the density matrices estimated from experimental projection measurements, for the single reference NW (presented in Fig. \ref{Fig1}b), (see Fig. \ref{Fif_S5}). The dataset of 16 measurements needed to estimate the density matrix is shown in Fig. \ref{Fif_S5b} for the H pump polarization. Owing to space limitations, only this particular case is shown but the dataset for every pump polarization is available upon reasonable request. 

\clearpage

\subsubsection{Quantum tomography for the two orthogonal NWs}
\begin{figure}[h!]
   \centering
    \includegraphics[width = 1\textwidth]{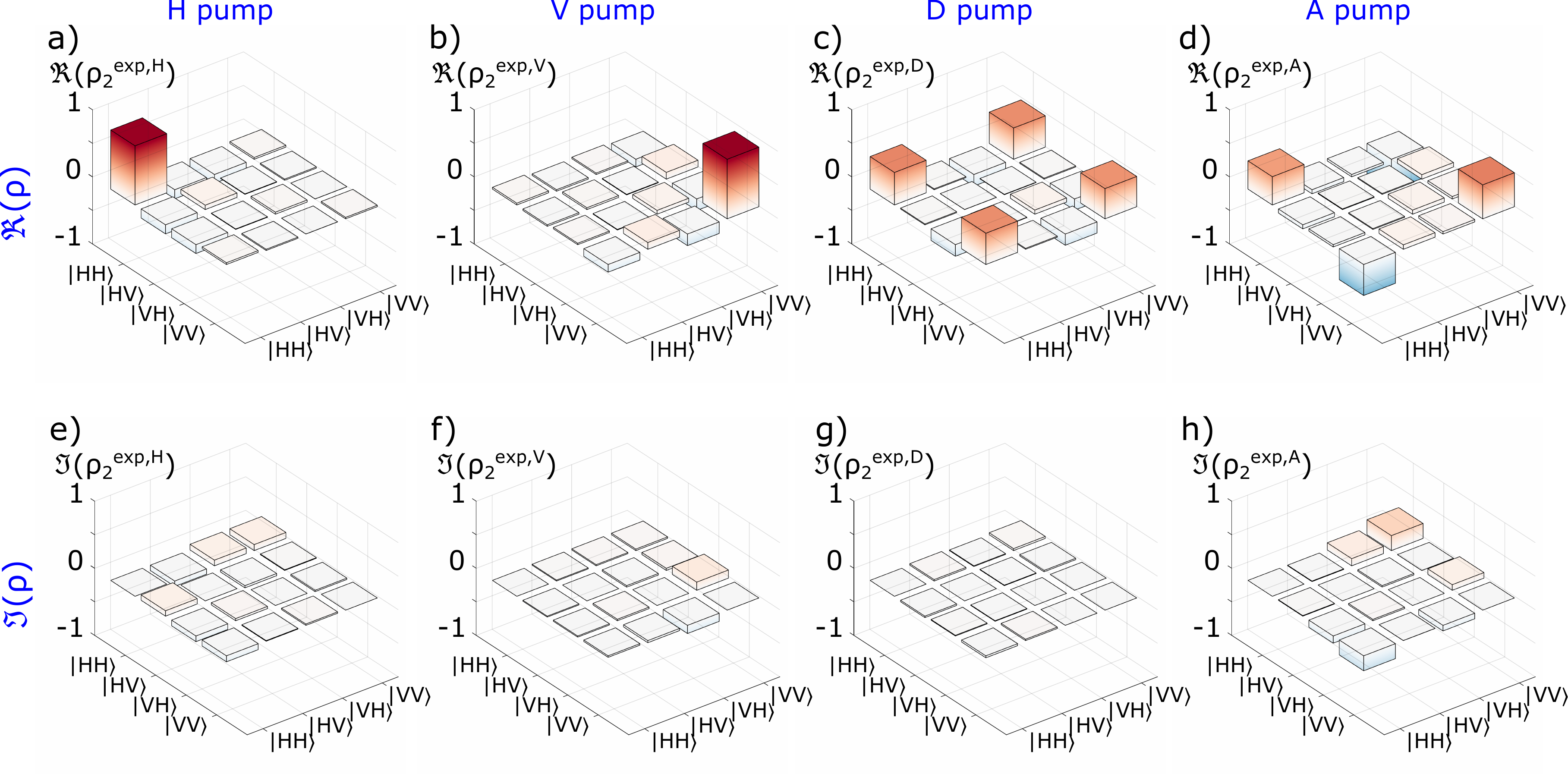}
  \caption{Experimental density matrice $\rho_{\mathrm{2}}^{\mathrm{exp,P}}$ measured for the two orthogonal NWs presented in Fig. \ref{Fig1}c. Each column correspond to a given P pump polarization being horizontal (H, with the polarization being aligned along the NW's long axis) for a),e); vertical (V) for b),f); diagonal (D) for c),g); and antidiagonal (A) for d),h). The real parts of the density matrices are shown in the first row (a-d), and the imaginary parts are shown in the second row (e-h). } 
  \label{Fig_S6}
\end{figure}

\begin{figure}[h!]
   \centering
    \includegraphics[width = 1\textwidth]{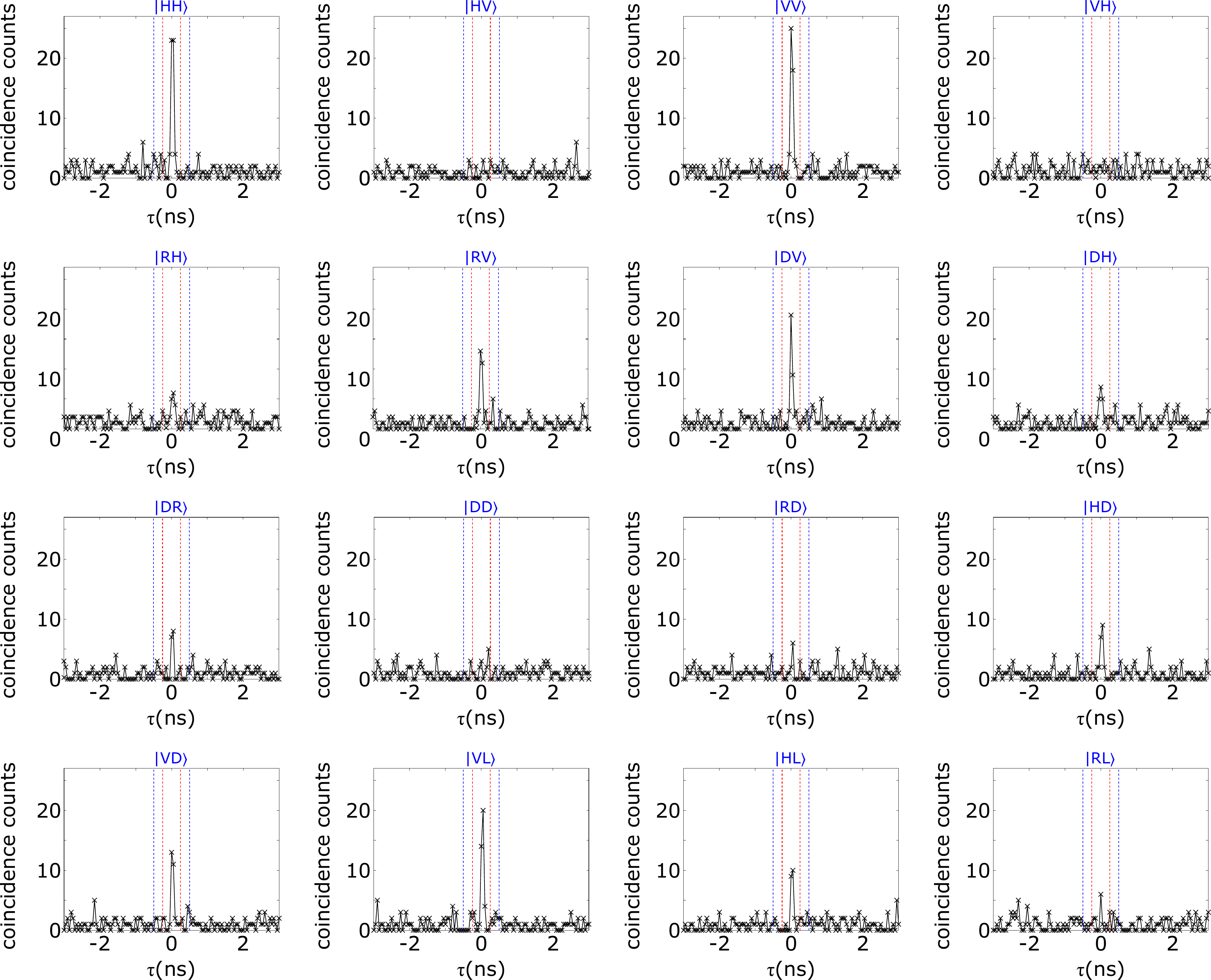}
  \caption{Experimental coincidence histograms of quantum state tomography for the reference single NW, for A pump polarization. The reconstruction of the density matrix shows a Bell state (see Fig. \ref{Fig5}h). The vertical red dashed lines indicate the coincidence window, while the blue dashed lines represent the area to exclude when calculating the background.} 
  \label{Fig_S6b}
\end{figure}

Similarly, we show in Fig. \ref{Fig_S6} the real and imaginary parts of the density matrices estimated from experimental projection measurements, for the two orthogonal NWs (presented in Fig. \ref{Fig1}c). The dataset of 16 measurements needed to estimate the density matrix is shown in Fig. \ref{Fig_S6b} for the A pump polarization. 

\clearpage

\begin{figure}[h!]
   \centering
    \includegraphics[width = 1\textwidth]{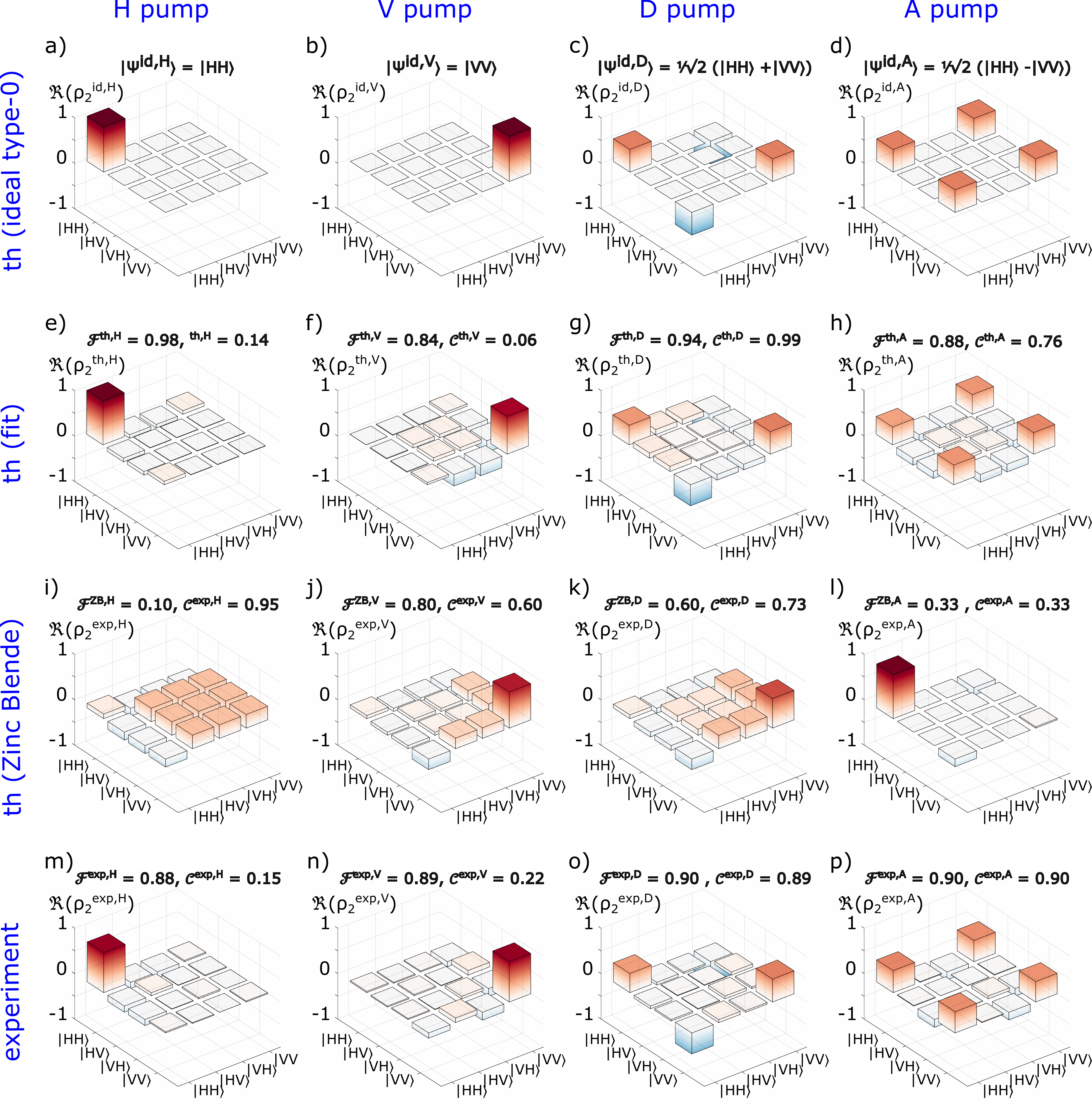}
  \caption{Polarization-entangled Bell state generation from the GaAs nanowires. a-d) Real part of the density matrices of an ideal type-0 process for which $\chi^{(2)}_{xxx} \neq 0$ and every other $\chi^{(2)}_{ijk}$ are zero.
  e-h) Real part of the theoretical density matrix $\rho_{\mathrm{2}}^{\mathrm{th,P}}$ estimated with the fit presented in the main manuscript, with P the pump polarization (See Eq. \ref{Eq4}). i-l) Real part of the theoretical density matrix estimated from the Zinc Blende susceptibility tensor (see Eq. \ref{Eq_ZB}). m-p) Real part of the experimental density matrix $\rho_{\mathrm{2}}^{\mathrm{exp,P}}$, with P the pump polarization. Each column correspond to a given P pump polarization being horizontal (H) for a),e),i),m); vertical (V) for b),f),j),n); diagonal (D) for c),g),k),o); and antidiagonal (A) for d),h),l),p). The fidelities $\mathcal{F}$ and concurrences $\mathcal{C}$ are computed for the comparison with ideal density matrices of a-d).} 
  \label{Fig_S7}
\end{figure}

The figure \ref{Fig_S7} displays the real part of the theoretical and experimental density matrices as a function of the P pump polarizations. \\

\begin{itemize}
    \item The first row of Fig. \ref{Fig_S7} shows the theoretical density matrices obtained for an ideal type-0 process, meaning $\chi^{(2)}_{xxx} \neq 0$ and every other $\chi^{(2)}_{ijk}$ are zero, leading to perfect Bell states $\ket{\Phi^{\pm}} = \dfrac{1}{\sqrt{2}}(\ket{HH}\pm\ket{VV})$ when pumped along D or A (see Fig. \ref{Fig_S7}c,d)).\\
    \item The second row shows the theoretical density matrices estimated from the effective susceptibility tensor presented in the main manuscript.\\
    \item The third row shows the theoretical density matrices estimated from the theoretical Zinc Blende susceptibility tensor \cite{saerens_background-free_2023}, which can be written as 
\begin{equation}
\chi^{(2)} _{\mathrm{ZB}}\propto
\begin{bmatrix}
1 & -0.5 & \mathsmaller{\mathrm{NA}} & \mathsmaller{\mathrm{NA}} & \mathsmaller{\mathrm{NA}} & 0 \\
0 & -0.71 & \mathsmaller{\mathrm{NA}} & \mathsmaller{\mathrm{NA}} & \mathsmaller{\mathrm{NA}} &-0.5 \\
\mathsmaller{\mathrm{NA}} & \mathsmaller{\mathrm{NA}} & \mathsmaller{\mathrm{NA}} & \mathsmaller{\mathrm{NA}} & \mathsmaller{\mathrm{NA}} & \mathsmaller{\mathrm{NA}} \\
\end{bmatrix}
\label{Eq_ZB}
\end{equation}
\noindent where $\mathsmaller{\mathrm{NA}}$ indicates data that cannot be accessed, as they are associated with the  z-polarization, perpendicular to the sample surface.\\
    \item The fourth row corresponds to the real parts of the experimental density matrices (see Fig. \ref{Fig_S6}, added to ease comparison with the theoretical ones).
\end{itemize}
 The fidelities displayed in Fig. \ref{Fig_S7} have been computed by comparing the density matrices with the ideal cases shown in Fig. \ref{Fig_S7}a-d).

\clearpage




\end{appendices}

\bibliography{biblio_entangled_NWs_nv}

@book{boyd2008nonlinear,
  title     = {Nonlinear Optics},
  author    = {Boyd, Robert W.},
  edition   = {3},
  year      = {2008},
  publisher = {Academic Press},
  address   = {Burlington, MA},
  isbn      = {978-0-12-369470-6},
}

@article{zhang_spatially_2022,
	title = {Spatially entangled photon pairs from lithium niobate nonlocal metasurfaces},
	volume = {8},
    year = {2022},
	issn = {2375-2548},
	url = {https://www.science.org/doi/10.1126/sciadv.abq4240},
	doi = {10.1126/sciadv.abq4240},
	pages = {eabq4240},
	journal = {Science Advances},
	shortjournal = {Sci. Adv.},
	author = {Zhang, Jihua and Ma, Jinyong and Parry, Matthew and Cai, Marcus and Camacho-Morales, Rocio and Xu, Lei and Neshev, Dragomir N. and Sukhorukov, Andrey A.},
	date = {2022-07-29},
	langid = {english},
}

@article{scarani_security_2009,
	title = {The security of practical quantum key distribution},
	volume = {81},
	rights = {http://link.aps.org/licenses/aps-default-license},
	issn = {0034-6861, 1539-0756},
	url = {https://link.aps.org/doi/10.1103/RevModPhys.81.1301},
	doi = {10.1103/RevModPhys.81.1301},
	pages = {1301--1350},
	journal = {Reviews of Modern Physics},
	shortjournal = {Rev. Mod. Phys.},
	author = {Scarani, Valerio and Bechmann-Pasquinucci, Helle and Cerf, Nicolas J. and Dušek, Miloslav and Lütkenhaus, Norbert and Peev, Momtchil},
	date = {2009-09-29},
    year = {2009},
	langid = {english},
}

@article{saerens_background-free_2023,
	title = {Background-Free Near-Infrared Biphoton Emission from Single {{G}a{A}s} Nanowires},
    year = {2023},
	volume = {23},
	rights = {https://creativecommons.org/licenses/by/4.0/},
	issn = {1530-6984, 1530-6992},
	url = {https://pubs.acs.org/doi/10.1021/acs.nanolett.3c00026},
	doi = {10.1021/acs.nanolett.3c00026},
	pages = {3245--3250},
	journal = {Nano Letters},
	shortjournal = {Nano Lett.},
	author = {Saerens, Grégoire and Dursap, Thomas and Hesner, Ian and Duong, Ngoc M. H. and Solntsev, Alexander S. and Morandi, Andrea and Maeder, Andreas and Karvounis, Artemios and Regreny, Philippe and Chapman, Robert J. and Danescu, Alexandre and Chauvin, Nicolas and Penuelas, José and Grange, Rachel},
	date = {2023-04-26},
	langid = {english},
}

@article{weissflog_directionally_2024,
	title = {Directionally Tunable Co- and Counter-Propagating Photon Pairs from a Nonlinear Metasurface},
	volume = {13},
    year = {2024},
	issn = {2192-8614},
	url = {http://arxiv.org/abs/2403.07651},
	doi = {10.1515/nanoph-2024-0122},
	pages = {3563--3573},
	journal = {Nanophotonics},
	author = {Weissflog, Maximilian A. and Ma, Jinyong and Zhang, Jihua and Fan, Tongmiao and Pertsch, Thomas and Neshev, Dragomir N. and Saravi, Sina and Setzpfandt, Frank and Sukhorukov, Andrey A.},
	date = {2024-08-20},
	langid = {english},
	eprinttype = {arxiv},
}

@article{weissflog_tunable_2024,
	title = {A tunable transition metal dichalcogenide entangled photon-pair source},
	volume = {15},
    year = {2024},
	issn = {2041-1723},
	url = {https://www.nature.com/articles/s41467-024-51843-3},
	doi = {10.1038/s41467-024-51843-3},
	pages = {7600},
	journal = {Nature Communications},
	shortjournal = {Nat Commun},
	author = {Weissflog, Maximilian A. and Fedotova, Anna and Tang, Yilin and Santos, Elkin A. and Laudert, Benjamin and Shinde, Saniya and Abtahi, Fatemeh and Afsharnia, Mina and Pérez Pérez, Inmaculada and Ritter, Sebastian and Qin, Hao and Janousek, Jiri and Shradha, Sai and Staude, Isabelle and Saravi, Sina and Pertsch, Thomas and Setzpfandt, Frank and Lu, Yuerui and Eilenberger, Falk},
	date = {2024-09-01},
	langid = {english},
}

@article{santiago-cruz_entangled_2021,
	title = {Entangled photons from subwavelength nonlinear films},
    year ={2021},
	volume = {46},
	issn = {0146-9592, 1539-4794},
	url = {https://opg.optica.org/abstract.cfm?URI=ol-46-3-653},
	doi = {10.1364/OL.411176},
	pages = {653-656},
	journal = {Optics Letters},
	shortjournal = {Opt. Lett.},
	author = {Santiago-Cruz, Tomás and Sultanov, Vitaliy and Zhang, Haizhong and Krivitsky, Leonid A. and Chekhova, Maria V.},
	date = {2021-02-01},
	langid = {english},
}

@article{defienne_polarization_2021,
	title = {Polarization entanglement-enabled quantum holography},
	volume = {17},
    year = {2021},
	issn = {1745-2473, 1745-2481},
	url = {https://www.nature.com/articles/s41567-020-01156-1},
	doi = {10.1038/s41567-020-01156-1},
	pages = {591--597},
	journal = {Nature Physics},
	shortjournal = {Nat. Phys.},
	author = {Defienne, Hugo and Ndagano, Bienvenu and Lyons, Ashley and Faccio, Daniele},
	date = {2021-05},
	langid = {english},
}

@article{santiago-cruz_photon_2021,
	title = {Photon Pairs from Resonant Metasurfaces},
	volume = {21},
    year = {2021},
	rights = {https://creativecommons.org/licenses/by/4.0/},
	issn = {1530-6984, 1530-6992},
	url = {https://pubs.acs.org/doi/10.1021/acs.nanolett.1c01125},
	doi = {10.1021/acs.nanolett.1c01125},
	pages = {4423--4429},
	journal = {Nano Letters},
	shortjournal = {Nano Lett.},
	author = {Santiago-Cruz, Tomás and Fedotova, Anna and Sultanov, Vitaliy and Weissflog, Maximilian A. and Arslan, Dennis and Younesi, Mohammadreza and Pertsch, Thomas and Staude, Isabelle and Setzpfandt, Frank and Chekhova, Maria},
	date = {2021-05-26},
	langid = {english},
}

@article{okoth_microscale_2019,
	title = {Microscale Generation of Entangled Photons without Momentum Conservation},
    year = {2019},
	volume = {123},
	issn = {0031-9007, 1079-7114},
	url = {https://link.aps.org/doi/10.1103/PhysRevLett.123.263602},
	doi = {10.1103/PhysRevLett.123.263602},
	pages = {263602},
	journal = {Physical Review Letters},
	shortjournal = {Phys. Rev. Lett.},
	author = {Okoth, C. and Cavanna, A. and Santiago-Cruz, T. and Chekhova, M.V.},
	date = {2019-12-31},
	langid = {english},
}

@article{okoth_idealized_2020,
	title = {Idealized Einstein-Podolsky-Rosen states from non–phase-matched parametric down-conversion},
	volume = {101},
    year ={2020},
	issn = {2469-9926, 2469-9934},
	url = {https://link.aps.org/doi/10.1103/PhysRevA.101.011801},
	doi = {10.1103/PhysRevA.101.011801},
	pages = {011801},
	journal = {Physical Review A},
	shortjournal = {Phys. Rev. A},
	author = {Okoth, C. and Kovlakov, E. and Bönsel, F. and Cavanna, A. and Straupe, S. and Kulik, S. P. and Chekhova, M. V.},
	date = {2020-01-14},
	langid = {english},
}

@article{james_measurement_2001,
	title = {Measurement of qubits},
    year = {2001},
	volume = {64},
	rights = {http://link.aps.org/licenses/aps-default-license},
	issn = {1050-2947, 1094-1622},
	url = {https://link.aps.org/doi/10.1103/PhysRevA.64.052312},
	doi = {10.1103/PhysRevA.64.052312},
	pages = {052312},
	journal = {Physical Review A},
	shortjournal = {Phys. Rev. A},
	author = {James, Daniel F. V. and Kwiat, Paul G. and Munro, William J. and White, Andrew G.},
	date = {2001-10-16},
	langid = {english},
}

@article{moreau_qimaging_2019,
	title = {Imaging with quantum states of light},
	volume = {1},
	issn = {2522-5820},
	url = {https://doi.org/10.1038/s42254-019-0056-0},
	doi = {10.1038/s42254-019-0056-0},
	pages = {367-380},
	journal = {Nature Reviews Physics},
	author = {Moreau, Paul-Antoine and Toninelli, Ermes and Gregory, Thomas and Padgett, Miles J.},
	date = {2019/06/01},
    year = {2019}
}

@article{fan_enhanced_2025,
	title = {Enhanced Photon-Pair Generation from a van der Waals Metasurface},
    year = {2025},
	rights = {https://doi.org/10.15223/policy-029},
	issn = {1530-6984, 1530-6992},
	url = {https://pubs.acs.org/doi/10.1021/acs.nanolett.5c02170},
	doi = {10.1021/acs.nanolett.5c02170},
	pages = {11844-11851},
    volume = {25},
	journal = {Nano Letters},
	shortjournal = {Nano Lett.},
	author = {Fan, Tongmiao and Tang, Yilin and Lung, Shaun and Weissflog, Maximilian and Ma, Jinyong and Shinde, Saniya and Saravi, Sina and Nauman, Mudassar and Yang, Wenkai and Qin, Hao and Qiu, Shuyao and Sukhorukov, Andrey A. and Lu, Yuerui and Setzpfandt, Frank},
	date = {2025-07-23},
	langid = {english},
}

@article{curty_entanglement_2004,
	title = {Entanglement as a Precondition for Secure Quantum Key Distribution},
	volume = {92},
	rights = {http://link.aps.org/licenses/aps-default-license},
	issn = {0031-9007, 1079-7114},
	url = {https://link.aps.org/doi/10.1103/PhysRevLett.92.217903},
	doi = {10.1103/PhysRevLett.92.217903},
	pages = {217903},
	journal = {Physical Review Letters},
	shortjournal = {Phys. Rev. Lett.},
	author = {Curty, Marcos and Lewenstein, Maciej and Lütkenhaus, Norbert},
	date = {2004-05-27},
    year  ={2004},
	langid = {english},
}

@article{gigli_quasinormal-mode_2020,
	title = {Quasinormal-Mode Non-Hermitian Modeling and Design in Nonlinear Nano-Optics},
	volume = {7},
    year = {2020},
	rights = {https://doi.org/10.15223/policy-029},
	issn = {2330-4022, 2330-4022},
	url = {https://pubs.acs.org/doi/10.1021/acsphotonics.0c00014},
	doi = {10.1021/acsphotonics.0c00014},
	pages = {1197--1205},
	journal = {{ACS} Photonics},
	shortjournal = {{ACS} Photonics},
	author = {Gigli, Carlo and Wu, Tong and Marino, Giuseppe and Borne, Adrien and Leo, Giuseppe and Lalanne, Philippe},
	date = {2020-05-20},
	langid = {english},
}

@article{sultanov_flat-optics_2022,
	title = {Flat-optics generation of broadband photon pairs with tunable polarization entanglement},
	volume = {47},
    year ={2022},
	issn = {0146-9592, 1539-4794},
	url = {https://opg.optica.org/abstract.cfm?URI=ol-47-15-3872},
	doi = {10.1364/OL.458133},
	pages = {3872-3875},
	journal = {Optics Letters},
	shortjournal = {Opt. Lett.},
	author = {Sultanov, Vitaliy and Santiago-Cruz, Tomás and Chekhova, Maria V.},
	date = {2022-08-01},
	langid = {english},
}

@article{jia_polarization-entangled_2025,
	title = {Polarization-entangled Bell state generation from an epsilon-near-zero metasurface},
	volume = {11},
	issn = {2375-2548},
	url = {https://www.science.org/doi/10.1126/sciadv.ads3576},
	doi = {10.1126/sciadv.ads3576},
	pages = {eads3576},
	journal = {Science Advances},
	shortjournal = {Sci. Adv.},
	author = {Jia, Wenhe and Saerens, Grégoire and Talts, Ulle-Linda and Weigand, Helena and Chapman, Robert J. and Li, Liu and Grange, Rachel and Yang, Yuanmu},
	date = {2025-02-21},
    year ={2025},
	langid = {english},
}

@article{noh_fano_2025,
	title = {Fano interference of photon pairs from a metasurface},
	doi = {https://doi.org/10.1038/s41377-025-01998-5},
	journal = {Light: Science \& Applications},
	author = {Noh, Jiho and Santiago-Cruz, Tomás and Doiron, Chloe and Jung, Hyunseung and Yu, Jaeyeon and Addamane, Sadhvikas and Chekhova, Maria and Brener, Igal},
    year = {2025},
    pages = {371},
    volume = {14},
}

@article{bouwmeester_experimental_1997,
	title = {Experimental quantum teleportation},
	author = {Bouwmeester, Dik and Pan, Jian-Wei and Mattle, Klaus and Eibl, Manfred and Weinfurter, Harald and Zeilinger, Anton},
	date = {1997},
	langid = {english},
    volume ={390},
    journal = {Nature},
    year = {1997},
    doi ={https://doi.org/10.1038/37539},
    pages ={575-579},
}

@article{erven_entangled_2008,
	title = {Entangled quantum key distribution over two free-space optical links},
	volume = {16},
	rights = {https://doi.org/10.1364/{OA}\_License\_v1\#{VOR}-{OA}},
	issn = {1094-4087},
	url = {https://opg.optica.org/oe/abstract.cfm?uri=oe-16-21-16840},
	doi = {10.1364/OE.16.016840},
	pages = {16840-16853},
	journal = {Optics Express},
	shortjournal = {Opt. Express},
	author = {Erven, C. and Couteau, C. and Laflamme, R. and Weihs, G.},
	date = {2008-10-13},
    year = {2008},
	langid = {english},
}

@article{liang_tunable_2025,
	title = {Tunable polarization entangled photon-pair source in rhombohedral boron nitride},
	volume = {11},
    year = {2025},
	issn = {2375-2548},
	url = {https://www.science.org/doi/10.1126/sciadv.adt3710},
	doi = {10.1126/sciadv.adt3710},
	pages = {eadt3710},
	journal = {Science Advances},
	shortjournal = {Sci. Adv.},
	author = {Liang, Haidong and Gu, Tian and Lou, Yanchao and Yang, Chengyuan and Ma, Chaojie and Qi, Jiajie and Bettiol, Andrew A. and Wang, Xilin},
	date = {2025-01-24},
	langid = {english},
}

@article{feng_polarization-entangled_2024,
	title = {Polarization-entangled photon-pair source with van der Waals 3R-{WS}2 crystal},
	volume = {4},
    year = {2023},
	issn = {2097-1710, 2662-8643},
	url = {https://elight.springeropen.com/articles/10.1186/s43593-024-00074-6},
	doi = {10.1186/s43593-024-00074-6},
	pages = {16},
	journal = {{eLight}},
	shortjournal = {{eLight}},
	author = {Feng, Jiangang and Wu, Yun-Kun and Duan, Ruihuan and Wang, Jun and Chen, Weijin and Qin, Jiazhang and Liu, Zheng and Guo, Guang-Can and Ren, Xi-Feng and Qiu, Cheng-Wei},
	date = {2024-12},
	langid = {english},
}

@article{abbas_recent_2024,
	title = {Recent progress, challenges, and opportunities in 2D materials for flexible displays},
	volume = {56},
    year = {2024},
	rights = {https://www.elsevier.com/tdm/userlicense/1.0/},
	issn = {1748-0132},
	url = {https://linkinghub.elsevier.com/retrieve/pii/S1748013224001117},
	doi = {10.1016/j.nantod.2024.102256},
	pages = {102256},
	journal = {Nano Today},
	author = {Abbas, Aumber and Luo, Yingjie and Ahmad, Waqas and Mustaqeem, Mujahid and Kong, Lingan and Chen, Jiwei and Zhou, Guigang and Tabish, Tanveer A. and Zhang, Qian and Liang, Qijie},
	date = {2024-06},
	langid = {english},
}

@article{guo_ultrathin_2023,
	title = {Ultrathin quantum light source with van der Waals {NbOCl}2 crystal},
    year = {2023},
	volume = {613},
	rights = {https://www.springernature.com/gp/researchers/text-and-data-mining},
	issn = {0028-0836, 1476-4687},
	url = {https://www.nature.com/articles/s41586-022-05393-7},
	doi = {10.1038/s41586-022-05393-7},
	pages = {53--59},
	journal = {Nature},
	author = {Guo, Qiangbing and Qi, Xiao-Zhuo and Zhang, Lishu and Gao, Meng and Hu, Sanlue and Zhou, Wenju and Zang, Wenjie and Zhao, Xiaoxu and Wang, Junyong and Yan, Bingmin and Xu, Mingquan and Wu, Yun-Kun and Eda, Goki and Xiao, Zewen and Yang, Shengyuan A. and Gou, Huiyang and Feng, Yuan Ping and Guo, Guang-Can and Zhou, Wu and Ren, Xi-Feng and Qiu, Cheng-Wei and Pennycook, Stephen J. and Wee, Andrew T. S.},
	date = {2023-01-05},
	langid = {english},
}

@article{kwiat_ultrabright_1999,
	title = {Ultrabright source of polarization-entangled photons},
	volume = {60},
	rights = {http://link.aps.org/licenses/aps-default-license},
	issn = {1050-2947, 1094-1622},
	url = {https://link.aps.org/doi/10.1103/PhysRevA.60.R773},
	doi = {10.1103/physreva.60.r773},
	pages = {R773--R776},
	journal = {Physical Review A},
	shortjournal = {Phys. Rev. A},
	author = {Kwiat, Paul G. and Waks, Edo and White, Andrew G. and Appelbaum, Ian and Eberhard, Philippe H.},
	date = {1999-08-01},
    year = {1999},
	langid = {english},
}

@article{kwiat_new_1995,
	title = {New High-Intensity Source of Polarization-Entangled Photon Pairs},
	volume = {75},
	rights = {http://link.aps.org/licenses/aps-default-license},
	issn = {0031-9007, 1079-7114},
	url = {https://link.aps.org/doi/10.1103/PhysRevLett.75.4337},
	doi = {10.1103/physrevlett.75.4337},
	pages = {4337--4341},
	journal = {Physical Review Letters},
	shortjournal = {Phys. Rev. Lett.},
	author = {Kwiat, Paul G. and Mattle, Klaus and Weinfurter, Harald and Zeilinger, Anton and Sergienko, Alexander V. and Shih, Yanhua},
	date = {1995-12-11},
    year = {1995},
	langid = {english},
}

@article{shoji_absolute_1997,
	title = {Absolute scale of second-order nonlinear-optical coefficients},
    year = {1997},
	volume = {14},
	rights = {https://doi.org/10.1364/{OA}\_License\_v1\#{VOR}},
	issn = {0740-3224, 1520-8540},
	url = {https://opg.optica.org/abstract.cfm?URI=josab-14-9-2268},
	doi = {10.1364/josab.14.002268},
	pages = {2268-2294},
	journal = {Journal of the Optical Society of America B},
	shortjournal = {J. Opt. Soc. Am. B},
	author = {Shoji, Ichiro and Kondo, Takashi and Kitamoto, Ayako and Shirane, Masayuki and Ito, Ryoichi},
	date = {1997-09-01},
	langid = {english},
}

@article{ma_polarization_2023,
	title = {Polarization Engineering of Entangled Photons from a Lithium Niobate Nonlinear Metasurface},
	volume = {23},
    year = {2023},
	rights = {https://doi.org/10.15223/policy-029},
	issn = {1530-6984, 1530-6992},
	url = {https://pubs.acs.org/doi/10.1021/acs.nanolett.3c02055},
	doi = {10.1021/acs.nanolett.3c02055},
	pages = {8091--8098},
	journal = {Nano Letters},
	shortjournal = {Nano Lett.},
	author = {Ma, Jinyong and Zhang, Jihua and Jiang, Yuxin and Fan, Tongmiao and Parry, Matthew and Neshev, Dragomir N. and Sukhorukov, Andrey A.},
	date = {2023-09-13},
	langid = {english},
}

@article{Pimenta16,
author = {Pimenta, A. C. S. and Teles Ferreira, D. C. and Roa, D. B. and Moreira, M. V. B. and de Oliveira, A. G. and González, J. C. and De Giorgi, M. and Sanvitto, D. and Matinaga, F. M.},
title = {Linear and Nonlinear Optical Properties of Single {G}a{A}s Nanowires with Polytypism},
journal = {The Journal of Physical Chemistry C},
volume = {120},
pages = {17046-17051},
year = {2016},
doi = {10.1021/acs.jpcc.6b04458},
URL = { https://doi.org/10.1021/acs.jpcc.6b04458},
}

@article{Timofeeva16,
author = {Timofeeva, Maria and Bouravleuv, Alexei and Cirlin, George and Shtrom, Igor and Soshnikov, Ilya and Reig Escal{\'e}, Marc and Sergeyev, Anton and Grange, Rachel},
title = {Polar Second-Harmonic Imaging to Resolve Pure and Mixed Crystal Phases along {G}a{A}s Nanowires},
journal = {Nano Letters},
volume = {16},
pages = {6290-6297},
year = {2016},
doi = {10.1021/acs.nanolett.6b02592},
URL = { https://doi.org/10.1021/acs.nanolett.6b02592},
}

@article{Guyot86,
  title = {General considerations on optical second-harmonic generation from surfaces and interfaces},
  author = {Guyot-Sionnest, P. and Chen, W. and Shen, Y. R.},
  journal = {Phys. Rev. B},
  volume = {33},
  issue = {12},
  pages = {8254--8263},
  numpages = {0},
  year = {1986},
  month = {Jun},
  publisher = {American Physical Society},
  doi = {10.1103/PhysRevB.33.8254},
  url = {https://link.aps.org/doi/10.1103/PhysRevB.33.8254}
}

@article{Zhang21,
author = {Zhang, Bin and Stehr, Jan E. and Chen, Ping-Ping and Wang, Xingjun and Ishikawa, Fumitaro and Chen, Weimin M. and Buyanova, Irina A.},
title = {Anomalously Strong Second-Harmonic Generation in {G}a{A}s Nanowires via Crystal-Structure Engineering},
journal = {Advanced Functional Materials},
volume = {31},
pages = {2104671},
doi = {https://doi.org/10.1002/adfm.202104671},
url = {https://advanced.onlinelibrary.wiley.com/doi/abs/10.1002/adfm.202104671},
year = {2021}
}

@article{Zhang13,
author = {Xiaoqing Zhang and Hao He and Jintao Fan and Chenglin Gu and Xin Yan and Minglie Hu and Xia Zhang and Xiaomin Ren and Chingyue Wang},
journal = {Opt. Express},
pages = {28432--28437},
publisher = {Optica Publishing Group},
title = {Sum frequency generation in pure zinc-blende {G}a{A}s nanowires},
volume = {21},
month = {Nov},
year = {2013},
url = {https://opg.optica.org/oe/abstract.cfm?URI=oe-21-23-28432},
doi = {10.1364/OE.21.028432},
}

@article{He13,
    author = {He, Hao and Zhang, Xiaoqing and Yan, Xin and Huang, Lili and Gu, Chenglin and Hu, Ming-lie and Zhang, Xia and Ren, Xiao min and Wang, Chingyue},
    title = {Broadband second harmonic generation in {G}a{A}s nanowires by femtosecond laser sources},
    journal = {Applied Physics Letters},
    volume = {103},
    pages = {143110},
    year = {2013},
    month = {10},
    issn = {0003-6951},
    doi = {10.1063/1.4824024},
    url = {https://doi.org/10.1063/1.4824024},
}

@article{Schaller02,
author = {Schaller, Richard D. and Johnson, Justin C. and Wilson, Kevin R. and Lee, Lynn F. and Haber, Louis H. and Saykally, Richard J.},
title = {Nonlinear Chemical Imaging Nanomicroscopy: From Second and Third Harmonic Generation to Multiplex (Broad-Bandwidth) Sum Frequency Generation Near-Field Scanning Optical Microscopy},
journal = {The Journal of Physical Chemistry B},
volume = {106},
pages = {5143-5154},
year = {2002},
doi = {10.1021/jp0144653},
URL = { https://doi.org/10.1021/jp0144653
},
}

@misc{Stich25,
      title={Thin-film Al0.30Ga0.70As (111) as a flat source of high-purity orthogonally polarized entangled photons}, 
      author={Simon Stich and Vitaliy Sultanov and Trevor Blakie and Qingyu Shi and Zbig Wasiliewski and Mikhail A. Belkin and Maria Chekhova},
      year={2025},
      eprint={2509.03978},
      archivePrefix={arXiv},
      primaryClass={physics.optics},
      url={https://arxiv.org/abs/2509.03978}, 
}

@article{Jozsa94,
author = {Richard Jozsa},
title = {Fidelity for Mixed Quantum States},
journal = {Journal of Modern Optics},
volume = {41},
pages = {2315--2323},
year = {1994},
publisher = {Taylor \& Francis},
doi = {10.1080/09500349414552171},
URL = {https://doi.org/10.1080/09500349414552171},
}

@article{Wootters98,
  title = {Entanglement of Formation of an Arbitrary State of Two Qubits},
  author = {Wootters, William K.},
  journal = {Phys. Rev. Lett.},
  volume = {80},
  issue = {10},
  pages = {2245--2248},
  numpages = {0},
  year = {1998},
  month = {Mar},
  publisher = {American Physical Society},
  doi = {10.1103/PhysRevLett.80.2245},
  url = {https://link.aps.org/doi/10.1103/PhysRevLett.80.2245}
}

@article{Modlawska08,
  title = {Nonmaximally Entangled States Can Be Better for Multiple Linear Optical Teleportation},
  author = {Mod\l{}awska, Joanna and Grudka, Andrzej},
  journal = {Phys. Rev. Lett.},
  volume = {100},
  issue = {11},
  pages = {110503},
  numpages = {4},
  year = {2008},
  month = {Mar},
  publisher = {American Physical Society},
  doi = {10.1103/PhysRevLett.100.110503},
  url = {https://link.aps.org/doi/10.1103/PhysRevLett.100.110503}
}

@article{Poddubny16,
  title = {Generation of Photon-Plasmon Quantum States in Nonlinear Hyperbolic Metamaterials},
  author = {Poddubny, Alexander N. and Iorsh, Ivan V. and Sukhorukov, Andrey A.},
  journal = {Phys. Rev. Lett.},
  volume = {117},
  issue = {12},
  pages = {123901},
  numpages = {6},
  year = {2016},
  month = {Sep},
  publisher = {American Physical Society},
  doi = {10.1103/PhysRevLett.117.123901},
  url = {https://link.aps.org/doi/10.1103/PhysRevLett.117.123901}
}

@Article{DeCeglia19,
AUTHOR = {de Ceglia, Domenico and Carletti, Luca and Vincenti, Maria Antonietta and De Angelis, Costantino and Scalora, Michael},
TITLE = {Second-Harmonic Generation in Mie-Resonant {G}a{A}s Nanowires},
JOURNAL = {Applied Sciences},
VOLUME = {9},
YEAR = {2019},
pages = {3381},
URL = {https://www.mdpi.com/2076-3417/9/16/3381},
ISSN = {2076-3417},
DOI = {10.3390/app9163381}
}

@Article{Guo24,
AUTHOR = {Guo, Qiangbing and Wu, Yun-Kun and Zhang, Di and Zhang, Qiuhong and Guo, Guang-Can and Alù, Andrea and Ren, Xi-Feng and Qiu, Cheng-Wei},
TITLE = {Polarization entanglement enabled by orthogonally stacked van der Waals NbOCl2 crystals},
JOURNAL = {Nature Communications},
VOLUME = {15},
YEAR = {2024},
ARTICLE-NUMBER = {2041-1723},
URL = {https://www.mdpi.com/2076-3417/9/16/3381},
DOI = {10.1038/s41467-024-54876-w},
	pages = {10461},
}

@article{Burnham70,
  title = {Observation of Simultaneity in Parametric Production of Optical Photon Pairs},
  author = {Burnham, David C. and Weinberg, Donald L.},
  journal = {Phys. Rev. Lett.},
  volume = {25},
  issue = {2},
  pages = {84--87},
  numpages = {0},
  year = {1970},
  month = {Jul},
  publisher = {American Physical Society},
  doi = {10.1103/PhysRevLett.25.84},
  url = {https://link.aps.org/doi/10.1103/PhysRevLett.25.84}
}

@article{Dursap25,
author = {Dursap, Thomas and Zhou, Tao and Dupraz, Maxime and Labat, Stéphane and Thomas, Olivier and Fardeau, Niels and Regreny, Philippe and Gendry, Michel and Brottet, Solène and Blanchard, Nicholas P. and Holt, Martin V. and Richard, Marie-Ingrid and Danescu, Alexandru and Penuelas, José and Bugnet, Matthieu},
title = {Correlated X-Ray and Electron Microscopies of a Single Biphasic {G}a{A}s Nanowire},
year = {2025},
journal = {Small Methods},
pages = {2500740},
doi = {https://doi.org/10.1002/smtd.202500740},
url = {https://onlinelibrary.wiley.com/doi/abs/10.1002/smtd.202500740}
}

@Article{Tabassom22,
AUTHOR = {Arjmand, Tabassom and Legallais, Maxime and Nguyen, Thi Thu Thuy and Serre, Pauline and Vallejo-Perez, Monica and Morisot, Fanny and Salem, Bassem and Ternon, Céline},
TITLE = {Functional Devices from Bottom-Up Silicon Nanowires: A Review},
JOURNAL = {Nanomaterials},
VOLUME = {12},
YEAR = {2022},
ARTICLE-NUMBER = {1043},
URL = {https://www.mdpi.com/2079-4991/12/7/1043},
PubMedID = {35407161},
ISSN = {2079-4991},
pages = {1043},
DOI = {10.3390/nano12071043}
}

@article{Hyun13,
   author = "Hyun, Jerome K. and Zhang, Shixiong and Lauhon, Lincoln J.",
   title = "Nanowire Heterostructures", 
   journal= "Annual Review of Materials Research",
   year = "2013",
   volume = "43",
   pages = "451-479",
   doi = "https://doi.org/10.1146/annurev-matsci-071312-121659",
   url = "https://www.annualreviews.org/content/journals/10.1146/annurev-matsci-071312-121659",
   publisher = "Annual Reviews",
   issn = "1545-4118",
   type = "Journal Article",
  }

@article{Edamatsu2007,
doi = {10.1143/JJAP.46.7175},
url = {https://doi.org/10.1143/JJAP.46.7175},
year = {2007},
month = {nov},
publisher = {},
volume = {46},
pages = {7175},
author = {Edamatsu, Keiichi},
title = {Entangled Photons: Generation, Observation, and Characterization},
journal = {Japanese Journal of Applied Physics}
}

@article{Lung20,
author = {Lung, Shaun and Wang, Kai and Kamali, Khosro Zangeneh and Zhang, Jihua and Rahmani, Mohsen and Neshev, Dragomir N. and Sukhorukov, Andrey A.},
title = {Complex-Birefringent Dielectric Metasurfaces for Arbitrary Polarization-Pair Transformations},
journal = {ACS Photonics},
volume = {7},
pages = {3015-3022},
year = {2020},
doi = {10.1021/acsphotonics.0c01044},
URL = { https://doi.org/10.1021/acsphotonics.0c01044},
}

@article{Kallioniemi25,
author = {Kallioniemi, Leevi andLyu, Xiaodan and He, Ruihua and  Rasmita, Abdullah and Duan, Ruihuan and Liu, Zheng and Gao, Weibo},
title = {Van der Waals engineering for quantum-entangled photon generation},
journal = {Nature Photonics},
volume = {19},
pages = {142-148},
year = {2025},
doi = {10.1038/s41566-024-01545-5},
URL = {https://doi.org/10.1038/s41566-024-01545-5},
}

@article{Dalacu19,
doi = {10.1088/1361-6528/ab0393},
url = {https://doi.org/10.1088/1361-6528/ab0393},
year = {2019},
month = {mar},
publisher = {IOP Publishing},
volume = {30},
pages = {232001},
author = {Dalacu, Dan and Poole, Philip J and Williams, Robin L},
title = {Nanowire-based sources of non-classical light},
journal = {Nanotechnology}
}

@article{Barrigon19,
author = {Barrig{ó}n, Enrique and Heurlin, Magnus and Bi, Zhaoxia and Monemar, Bo and Samuelson, Lars},
title = {Synthesis and Applications of III–V Nanowires},
journal = {Chemical Reviews},
volume = {119},
pages = {9170-9220},
year = {2019},
doi = {10.1021/acs.chemrev.9b00075},
URL = { https://doi.org/10.1021/acs.chemrev.9b00075}}

@article{Gisin07,
author = {Gisin, Nicolas and Thew, Rob},
title = {Quantum communication},
journal = {Nature Photonics},
volume = {1},
pages = {165-171},
year = {2007},
doi = {10.1038/nphoton.2007.22},
URL = { https://doi.org/10.1038/nphoton.2007.22},
}

@article{Pittaluga25,
author = {Pittaluga, Mirko and Lo, Yuen San and Brzosko, Adam and Woodward, Robert I. and Scalcon, Davide and Winnel, Matthew S. and Roger, Thomas and Dynes, James F. and Owen, Kim A. and Juárez, Sergio and Rydlichowski, Piotr and Vicinanza, Domenico and Roberts, Guy and Shields, Andrew J.},
title = {Long-distance coherent quantum communications in deployed telecom networks},
journal = {Nature},
volume = {640},
pages = {911-917},
year = {2025},
doi = {10.1038/s41586-025-08801-w},
URL = {https://doi.org/10.1038/s41586-025-08801-w},
}

@article{Conache08,
doi = {10.1088/1742-6596/100/5/052051},
url = {https://doi.org/10.1088/1742-6596/100/5/052051},
year = {2008},
month = {mar},
publisher = {},
volume = {100},
pages = {052051},
author = {G Conache and S Gray and M Bordag and A Ribayrol and L E Fröberg and L Samuelson and H Pettersson and L Montelius},
title = {AFM-based manipulation of InAs nanowires},
journal = {Journal of Physics: Conference Series}
}

@article{Junno95,
    author = {Junno, T. and Deppert, K. and Montelius, L. and Samuelson, L.},
    title = {Controlled manipulation of nanoparticles with an atomic force microscope},
    journal = {Applied Physics Letters},
    volume = {66},
    pages = {3627-3629},
    year = {1995},
    month = {06},
    issn = {0003-6951},
    doi = {10.1063/1.113809},
    url = {https://doi.org/10.1063/1.113809},
}

\end{document}